\documentclass[preprintnumbers,amsmath,amssymb,onecolumn]{revtex4}%

\usepackage{multirow}
\usepackage{graphicx}
\usepackage{dcolumn}
\usepackage{bm}
\usepackage{color}
\usepackage{lmodern}
\usepackage{longtable}

\usepackage{hyperref}
\hypersetup{
  colorlinks=true,        
  linkcolor=blue,         
  citecolor=magenta,         
}

\begin{document}

\title{\boldmath Rotating black hole solutions with quintessential energy}

\author{Bobir Toshmatov$^1$}
\email{b.a.toshmatov@gmail.com}

\author{Zden\v{e}k Stuchl\'{i}k$^1$}
\email{zdenek.stuchlik@fpf.slu.cz}

\author{Bobomurat Ahmedov$^{2,3}$}
\email{ahmedov@astrin.uz}

\affiliation{$^{1}$ Institute of Physics and Research Centre of Theoretical Physics and Astrophysics, Faculty of Philosophy \& Science, Silesian University in Opava, Bezru\v{c}ovo n\'{a}m\v{e}st\'{i} 13, CZ-74601 Opava, Czech Republic\\
$^{2}$ Institute of Nuclear Physics, Ulughbek, Tashkent 100214, Uzbekistan\\
$^{3}$ Ulugh Beg Astronomical Institute, Astronomicheskaya 33, Tashkent 100052, Uzbekistan}

\begin{abstract}
Quintessential dark energy with density $\rho$ and pressure $p$ is governed by an equation of state of the form $p=-\omega_{q}\rho$ with the quintessential parameter $\omega_q\in(-1;-1/3)$. We derive the geometry of quintessential rotating black holes, generalizing thus the Kerr spacetimes. Then we study the quintessential rotating black hole spacetimes with the special value of $\omega_q = -2/3$ when the resulting formulae are simple and easily tractable. We show that such special spacetimes can exist for dimensionless quintessential parameter $c<1/6$ and determine the critical rotational parameter $a_0$ separating the black hole and naked singularity spacetime in dependence on the quintessential parameter $c$. For the spacetimes with $\omega_q = 2/3$ we present the integrated geodesic equations in separated form and study in details the circular geodetical orbits. We give radii and parameters of the photon circular orbits, marginally bound and marginally stable orbits. We stress that the outer boundary on the existence of circular geodesics, given by the so called static radius where the gravitational attraction of the black hole is balanced by the cosmic repulsion, does not depend on the dimensionless spin of the rotating black hole, similarly to the case of the Kerr-de Sitter spacetimes with vacuum dark energy. We also give restrictions on the dimensionless parameters $c$ and $a$ of the spacetimes allowing for existence of stable circular geodesics.
\end{abstract}

\maketitle

\section{Introduction}\label{sec:intro}

Cosmological observations indicate that in the inflationary paradigm~\cite{Lin:1990:InfCos:}, a very small relict repulsive cosmological constant $\Lambda > 0$, i.e., vacuum energy, or, generally, a dark energy demonstrating repulsive gravitational effect, has to be invoked to explain the dynamics of the recent Universe~\cite{Kra-Tur:1995:GENRG2:,Ost-Ste:1995:NATURE:,Kra:1998:ASTRJ2:,Bah-etal:1999:SCIEN:, Cal-Dav-Ste:1998:PHYRL:,ArP-Muk-Ste:2000:PHYRL:,Wan-etal:2000:ASTRJ2:}. These results are confirmed by recent measurements of cosmic microwave background anisotropies obtained by the space satellite observatory PLANCK~\cite{Ade-etal:2014:ASTRA:}. Moreover, observations of distant Ia-type supernova explosions indicate that starting at the cosmological redshift $z \approx 1$ expansion of the Universe is accelerated~\cite{Rie-etal:2004:ASTRJ2:}.

The cosmological tests demonstrate that the dark energy represents about $70\%$ of the energy content of the observable universe~\cite{Spe-etal:2007:ApJSuppl:,Cal-Kam:2009:NATURE:CosDarkMat}. The dark energy equation of state is close to those corresponding to the vacuum energy, represented by the repulsive cosmological constant, but a different form of dark energy related to the so called quintessence is not excluded~\cite{Cal-Kam:2009:NATURE:CosDarkMat}. The cosmological and astrophysical effects of the cosmic repulsion indicate recent value of the cosmological constant to be $\Lambda \approx 1.3\times 10^{-56}\,\mathrm{cm^{-2}}$ \cite{Stu:2005:MPLA:}.

The presence of the repulsive cosmological constant changes dramatically the asymptotic structure of black-hole, naked singularity, or any compact-body backgrounds as such backgrounds become asymptotically de Sitter spacetimes, not flat spacetimes. In such spacetimes, an event horizon (cosmological horizon) always exists, behind which the geometry is dynamic. The cosmological horizon exists also in the spacetimes of black holes or other compact objects immersed in a quintessential field.

Here we first derive the geometry of spacetime describing rotating black hole with quintessential fields. Then we discuss in detail the circular geodesics of the rotating black holes with quintessential parameters $\omega_q=-2/3$ when the resulting formula can be given in a relatively simple form enabling an intuitive insight into accretion processes.

The repulsive cosmological constant was first discussed in the cosmological models~\cite{Mis-Tho-Whe:1973:Gra:}. Its role in the vacuola models of mass concentrations immersed in the expanding universe has been considered in~\cite{Stu:1983:BULAI:,Stu:1984:BULAI:,Far-etal:2009:PHYSLETB:,Gre-Lak:2010:PHYSR4:,Uza-Ell-Lar:2011:GENRG2:, Gre-Lak:2011:PHYSR4:,Fle-Dup-Uza:2013:PHYSR4:,Far-etal:2014:PHYSR4:,Far:2015:JCAP:,Far-etal:2015:JCAP:} and in the McVittie model \cite{McV:1933:MONRAS:} of mass concentrations immersed in the expanding universe \cite{Nol:1998:PHYSR4,Nol:1999:CLAQG:,Nan-Las-Hob:2012:MONRAS:,Kal-Kle-Mar:2010:PHYSR4:,Lak-Abd:2011:PHYSR4:,Sil-Fon-Gua:2013:PHYSR4:,Nol:2014:CLAQG:}.

Significant role of the repulsive cosmological constant has been demonstrated also for astrophysical situations related to active galactic nuclei and their central supermassive black holes~\cite{Stu:2005:MODPLA:}. The black hole spacetimes with the $\Lambda$ term are described in the spherically symmetric case by the vacuum Schwarzschild-de Sitter (SdS) geometry \cite{Kot:1918:ANNPH2:PhyBasEinsGr,Stu-Hle:1999:PHYSR4:}, while the internal, uniform density SdS spacetimes are given in~\cite{Stu:2000:ACTPS2:,Boh:2004:GENRG2:}). In axially symmetric, rotating case, the vacuum spacetime is determined by the Kerr-de Sitter (KdS) geometry \cite{Car:1973:BlaHol:}. In the spacetimes with the repulsive cosmological term, motion of photons is treated in a series of papers~\cite{Stu-Cal:1991:GENRG2:,Stu-Hle:2000:CLAQG:,Lak:2002:PHYSR4:,Bak-etal:2007:CEURJP:, Ser:2008:PHYSR4:,Sch-Zai:2008:0801.3776:CCTimeDelay,Mul:2008:GENRG2:FallSchBH,Vil-etal:2013:ASTSS1:} while motion of test particles was studied in~\cite{Stu:1983:BULAI:,Stu-Hle:1999:PHYSR4:,Stu-Sla:2004:PHYSR4:,Kra:2004:CLAQG:, Kra:2005:DARK:CCPerPrec,Kra:2007:CLAQG:Periapsis,Kag-Kun-Lam:2006:PHYLB:SolarSdS, Ali:2007:PHYSR4:EMPropKadS,Ior:2009:NEWASTR:CCDGPGrav,
Hac-etal:2010:PHYSR4:KerrBHCoStr,Oli-etal:2011:MODPLA:ChaParRNadS}. Oscillatory motion of current carrying string loops in SdS and KdS spacetimes was treated in~\cite{Kol-Stu:2010:PHYSR4:,Stu-Kol:2012:PHYSR4:,Gu-Cheng:2007:GENRG2:,Wan-Che:2012:PHYLB:}.

The cosmological constant can be relevant in both the geometrically thin Keplerian accretion discs \cite{Stu:2005:MODPLA:,Stu-Hle:1999:PHYSR4:,Stu-Sla:2004:PHYSR4:,Mul-Asch:2007:CLAQG:,Sla-Stu:2008:CLAQG:} and geometrically thick toroidal accretion discs~\cite{Stu-Sla-Hle:2000:ASTRA:,Sla-Stu:2005:CLAQG:,Rez-Zan-Fon:2003:ASTRA:,Stu-etal:2005:PHYSR4:,Mul-Asch:2007:CLAQG:,Asc:2008:CHIAA:, Sla-Stu:2008:CLAQG:,Kuc-Sla-Stu:2011:JCAP:} orbiting supermassive black holes in the central parts of giant galaxies. In spherically symmetric spacetimes the Keplerian and toroidal disc structures can be described by an appropriately chosen Pseudo-Newtonian potential~\cite{Stu-Kov:2008:INTJMD:,Stu-Sla-Kov:2009:CLAQG:} that is useful also in studies of the motion of interacting galaxies~\cite{Stu-Sch:2011:JCAP:,Sche-Stu-Pet:2013:JCAP:,Stu-Sch:2012:INTJMD:}. The KdS geometry can be relevant also in the case of Kerr superspinars representing an alternate explanation of active galactic nuclei~\cite{Gim-Hor:2004:hep-th0405019:GodHolo,Gim-Hor:2009:PHYLB:,Stu-Sch:2012:CLAQG:} -- the superspinars break the black hole bound on the spin and exhibit a variety of extraordinary physical phenomena~\cite{deFel:1974:ASTRA:,deFel:1978:NATURE:,Stu:1980:BULAI:,Stu:1981:BULAI:,Hio-Mae:2009:PHYSR4:, Stu-Sch:2010:CLAQG:,Stu-Sch:2012:CLAQG:,Stu-Sch:2013:CLAQG:}. The cosmological constant is also relevant in the internal structure of extended astrophysical objects governed by polytropic equations of state that could represent galactic dark matter halos or larger structures \cite{Stu-Sch:2012:CLAQG:}.

The spherically symmetric solutions describing a black hole surrounded by dark energy in the form of a quintessential field with equation of state in the form $p=\omega_{q}\rho$, with the quintessential parameter $-1<\omega_q<-1/3$, has been found by Kiselev \cite{Kis:2002:CLAQG:}. Of course, it is relevant to study the geodesic structure of such spacetime, and its influence in astrophysical phenomena, e.g., the accretion discs structure. The geodesic structure of the spherically symmetric quintessential black hole spacetimes has been studied in \cite{Fer-etal:2003:IJTP:}. Here we first generalize the Kiselev solution to the case of the rotating black holes for general value of the quintessential parameter $\omega_q$. Then we discuss in detail the circular geodesic orbits around quintessential rotating black holes with a special value of the quintessential parameter $\omega_q=-2/3$ when the results can be expressed in a significantly simple form, especially in the case of spherically symmetric spacetime. The other simple case of the dark energy is represented by the vacuum energy related to the cosmological constant when $\omega_q=1$.

\section{Quintessential rotating black hole solution}

By using the standard Newman-Janis algorithm (NJA) of generating rotating black hole spacetimes from spherically symmetric black hole spacetimes, we find quintessential rotating black hole spacetimes.

\subsection{Schwarzschild black hole surrounded by the quintessential energy}

Spacetime of the spherically symmetric quintessential black hole is determined by the line element
\begin{eqnarray}\label{01}
ds^2=-f(r)dt^2+g^{-1}(r)dr^2+h(r) d\Omega^2
\end{eqnarray}
where $h(r)=r^2$ and the lapse function $f(r)$ is given by the expression~\cite{Kis:2002:CLAQG:}
\begin{eqnarray}\label{02}
f(r)=g(r)=1-\frac{2M}{r}-\frac{c}{r^{3\omega_q+1}}.
\end{eqnarray}
In this expression, $M$ is the gravitational mass of the black hole and $c$ is the quintessential parameter representing intensity of the quintessence energy field related to the black hole. Dimensionless quintessential EoS (equation of state) parameter $\omega_q$ governs the equation of state of the quintessential field $p=\omega_{q}\rho$, relating pressure $p$ and energy density $\rho$ of the quintessential field, and it can vary in the range $\omega_q\in(-1;-1/3)$. Due to the quintessential field, the geometry is not Ricci flat. Here and in the following we use the geometric units with speed of light and Newtonian gravitational constant are taken as 1 $c=G=1$.

One can see from the lapse function~(\ref{02}) that for vanishing of the quintessential field ($c=0$) we recover the Schwarzschild black hole spacetime. On the other hand, for the EoS quintessential parameter $\omega_q=-1$ we obtain a vacuum energy equation of state $p=-\rho$, and the metric describes Schwarzschild-de-Sitter black holes.

Due to the quintessential field the geometry is not Ricci flat. It is well known that the singularities appeared in the metric by setting $g_{tt}=0$ and $g^{rr}=0$ are coordinate singularities which can be removed by using more appropriate set of coordinates. The coordinate singularities of the quintessential Kiselev black hole spacetime determine the black hole and the cosmological horizons given by the condition
\begin{eqnarray}\
f(r,M,c,\omega_c) = 0 .
\end{eqnarray}

One way to check the existence of physical singularity is to calculate curvature scalars, such as the Kretschmann scalar $K$ giving a square of the Riemann tensor. By calculating this we find
\begin{eqnarray}\label{03}
K = R^{\mu\nu\varrho\sigma}R_{\mu\nu\varrho\sigma}=\frac{W(r)}{r^{6\omega_q+6}}\ ,
\end{eqnarray}
where $W(r)$ is some polynomial function of $r$ ($W(r=0)\neq0$). Therefore one can conclude that a physical singularity occurs at $r=0$.

\subsection{Generating quintessential rotating black hole solutions}

Using the standard method of generating a rotational spacetime from a spherically symmetric spacetime proposed by Newman and Janis \cite{New-Jan:1965:JMP:}, and modifications introduced in works of Azreg-Ainou~\cite{Azreg-Ainou2014PRD,Azreg-Ainou11,Azreg-Ainou14}, we convert static spherically symmetric quintessential BH solution into the rotational form. The only difference of this method from the NJA method is skipping one of steps of the NJA, namely, the complexification of coordinates. Instead of this we perform "alternate" coordinate transformation.

At the first step of this algorithm we turn the spacetime metric~(\ref{01}) from the Boyer-Lindquist (BL) coordinates ($t,r,\theta,\phi$) to the Eddington-Finkelstein (EF) coordinates ($u,r,\theta,\phi$) by introducing the following coordinate transformations:
\begin{eqnarray}\label{2.1}
du=dt-\frac{dr}{\sqrt{fg}}\ .
\end{eqnarray}
As the result of this transformation we obtain the spacetime metric in the form
\begin{eqnarray}\label{2.2}
ds^2=-fdu^2-2\sqrt{\frac{f}{g}}dudr+h^2d\theta^2+h\sin^2\theta d\phi^2\ .
\end{eqnarray}
It is known that contravariant components of the metric tensor in the advanced null EF coordinates can be expressed by the null tetrad taking the form
\begin{eqnarray}\label{2.3}
g^{\mu\nu}=-l^\mu n^\nu-l^\nu n^\mu+m^\mu\bar{m}^\nu+m^\nu\bar{m}^\mu\ ,
\end{eqnarray}
with
\begin{eqnarray}\label{2.4}
&&l^\mu=\delta_r^\mu\ , \quad n^\mu=\sqrt{\frac{g}{f}}\delta_u^\mu-\frac{f}{2}\delta_r^\mu\ ,\nonumber\\
&&m^\mu=\frac{1}{\sqrt{2h}}\delta_\theta^\mu+\frac{i}{\sqrt{2h}\sin\theta}\delta_\phi^\mu\ , \quad \bar{m}^\mu=\frac{1}{\sqrt{2h}}\delta_\theta^\mu-\frac{i}{\sqrt{2h}\sin\theta}\delta_\phi^\mu\ .
\end{eqnarray}
Vectors $l$ and $n$ are real, $m$ is a complex vector, $\bar{m}$ vector is a complex conjugate of the vector $m$. These vectors satisfy orthogonality $l^\mu m_\mu=l^\mu \bar{m}_\mu=n^\mu m_\mu=n^\mu \bar{m}_\mu=0$, isotropic $l^\mu l_\mu=n^\mu n_\mu=m^\mu m_\mu=\bar{m}^\mu \bar{m}_\mu=0$, and normalization $l^\mu n_\mu=1$, $m^\mu\bar{m}_\mu=-1$ conditions.

Now we perform complex coordinate transformations in the $u-r$ plane
\begin{eqnarray}\label{2.5}
u\rightarrow u-ia\cos\theta, \quad r\rightarrow r-ia\cos\theta .
\end{eqnarray}
We assume that as the result of these transformations the metric functions also turn into a new form: $f(r)\rightarrow F(r,a,\theta)$, $g(r)\rightarrow G(r,a,\theta)$, $h(r)\rightarrow \Sigma(r,a,\theta)$. In the case $a=0$ new functions reduce to initial forms. Furthermore, null tetrads also take the new form
\begin{eqnarray}\label{2.6}
&&l^\mu=\delta_r^\mu\ , \quad m^\mu=\frac{1}{\sqrt{2\Sigma}}\left[\delta_\theta^\mu+ia\sin\theta(\delta_u^\mu-\delta_r^\mu)+\frac{i}{\sin\theta}\delta_\phi^\mu\right]\ ,\nonumber\\
&&n^\mu=\sqrt{\frac{G}{F}}\delta_u^\mu-\frac{1}{2}F\delta_r^\mu\ , \quad \bar{m}^\mu=\frac{1}{\sqrt{2\Sigma}}\left[\delta_\theta^\mu-ia\sin\theta(\delta_u^\mu-\delta_r^\mu)-\frac{i}{\sin\theta}\delta_\phi^\mu\right]\ .
\end{eqnarray}
Then we can rewrite the contravariant components of the metric tensor $g^{\mu\nu}$ by using~(\ref{2.3}) as
\begin{eqnarray} \label{2.7}
g^{\mu\nu} =      \left(
\begin{array}{c c c c}
a^2\sin^2\theta/\Sigma & -\sqrt{G/F}-a^2\sin^2\theta/\Sigma & 0 &a/\Sigma \\
-\sqrt{G/F}-a^2\sin^2\theta/\Sigma & G+a^2\sin^2\theta/\Sigma & 0 & -a/\Sigma \\
0 & 0 & 1/\Sigma & 0 \\
a/\Sigma & -a/\Sigma & 0 & 1/(\Sigma\sin^2\theta) \\
\end{array} \right).
\end{eqnarray}
The covariant components read
\begin{eqnarray} \label{2.8}
g_{\mu\nu} =      \left(
\begin{array}{c c c c}
-F & -\sqrt{F/G} & 0 &a\left(F-\sqrt{F/G}\right)\sin^2\theta \\
-\sqrt{F/G} & 0 & 0 & a\sqrt{F/G}\sin^2\theta \\
0 & 0 & \Sigma & 0 \\
a\left(F-\sqrt{F/G}\right)\sin^2\theta & a\sqrt{F/G}\sin^2\theta & 0 & \sin^2\theta\left[\Sigma+a^2\left(2\sqrt{F/G}-F\right)\sin^2\theta\right] \\
\end{array} \right).
\end{eqnarray}
The last step of the algorithm is the turn back from the EF coordinates to the BL coordinates by using the following coordinate transformations:
\begin{eqnarray}\label{2.9}
du=dt+\lambda(r)dr\ , \quad d\phi=d\phi+\chi(r)dr\ .
\end{eqnarray}
The transformation functions $\lambda(r)$ and $\chi(r)$ are found due to the requirement that all the non-diagonal components of the metric tensor, except the coefficient $g_{t\phi}$ ($g_{\phi t}$), are equal to zero~\cite{Azreg-Ainou2014PRD}. Thus
\begin{eqnarray}\label{2.10}
\lambda(r)=-\frac{k(r)+a^2}{g(r) h(r)+a^2}\ ,\quad \chi(r)=-\frac{a}{g(r) h(r)+a^2}
\end{eqnarray}
where
\begin{eqnarray}\label{2.11}
k(r)=\sqrt{\frac{g(r)}{f(r)}}h(r)\ ,
\end{eqnarray}
and
\begin{eqnarray}\label{2.12}
F(r,\theta)=\frac{(gh+a^2\cos^2\theta)\Sigma}{(k+a^2\cos^2\theta)^2},\quad G(r,\theta)=\frac{gh+a^2\cos^2\theta} {\Sigma} . \nonumber\\
\end{eqnarray}
Finally we obtain the line element of the rotational version of the quintessential Schwarzscild spacetime~(\ref{01}) in the form
\begin{eqnarray}\label{2.13}
ds^2=-\frac{(gh+a^2\cos^2\theta)\Sigma}{(k+a^2\cos^2\theta)^2}dt^2&+&\frac{\Sigma}{gh+a^2}dr^2 -2a\sin^2\theta\left[\frac{k-gh}{(k+a^2\cos^2\theta)^2}\right]\Sigma d\phi dt+\Sigma dt^2 \nonumber\\ &&+\Sigma\sin^2\theta\left[1+a^2\sin^2\theta\frac{2k-gh+a^2\cos^2\theta}{(k+a^2\cos^2\theta)^2}\right]d\phi^2 .
\end{eqnarray}
In our case $f(r)=g(r)$ and $h(r)=r^2$~\cite{Kis:2002:CLAQG:}. In this case there is $k(r)=h(r)$ and comparing the functions~(\ref{2.12}), we find
\begin{eqnarray}\label{2.14}
\Sigma=r^2+a^2\cos^2\theta\ .
\end{eqnarray}
Then the rotational version of the Kiselev quintessential solution takes the form
\begin{eqnarray}\label{2.15}
ds^2&=&-\left(1-\frac{2Mr+c r^{1-3\omega_q}}{\Sigma}\right)dt^2+\frac{\Sigma}{\Delta}dr^2-2a\sin^2\theta\left(\frac{2Mr+c r^{1-3\omega_q}}{\Sigma}\right)d\phi dt+\Sigma d\theta^2 \nonumber\\&&+\sin^2\theta\left[r^2+a^2+a^2\sin^2\theta\left(\frac{2Mr+c r^{1-3\omega_q}}{\Sigma}\right)\right]d\phi^2\ ,
\end{eqnarray}
where
\begin{eqnarray}\label{2.14}
\Delta=r^2-2Mr+a^2-c r^{1-3\omega_q} .
\end{eqnarray}
In the case of vanishing quintessential field, $c=0$, the spacetime metric~(\ref{2.15}) coincides with the Kerr one.

\section{Rotating black hole spacetimes with the parameter $\omega_{q}=-2/3$}

Now we focus our attention on the special case of the quintessential field with the parameter $\omega=-2/3$ enabling a relatively simple treatment of the spacetime properties.

\subsection{The geometry and the quintessential field}

Introducing a new notation and putting $\omega_{q}=-2/3$ we rewrite the spacetime metric (\ref{2.15}) in the Kerr-like form
\begin{eqnarray}\label{2.16}
ds^2=-\left(1-\frac{2\rho r}{\Sigma}\right)dt^2+\frac{\Sigma}{\Delta}dr^2-\frac{4a\rho r\sin^2\theta}{\Sigma}d\phi dt+\Sigma d\theta^2 +\sin^2\theta\left(r^2+a^2+a^2\sin^2\theta\frac{2\rho r}{\Sigma}\right)d\phi^2\ ,
\end{eqnarray}
where
\begin{eqnarray}\label{2.17}
\Delta(r)=r^2-2\rho r+a^2, \quad 2\rho(r)=2M+cr^2 .
\end{eqnarray}

From the quintessential rotational black hole spacetime metric and the Einstein equations $G_{\mu\nu} = 8\pi T_{\mu\nu}$, we find the quintessential stress-energy tensor assumed in the form $T_{\mu\nu} = (\epsilon,p_r,p_\theta,p_\phi)$ \cite{Kis:2002:CLAQG:}.
Using the Mathematica package RGTC \cite{RGTC} we find from the metric (\ref{2.16.1}) the nonzero components of the Einstein tensor $G_{\mu\nu}$ in the form
\begin{eqnarray}\label{Einstein}
&&G_{tt}=\frac{2\left[r^4-2r^3\rho+a^2r^2-a^4\sin^2\theta\cos^2\theta\right]\rho'}{\Sigma^3}-\frac{a^2r\sin^2\theta \rho''}{\Sigma^2},\nonumber\\
&&G_{rr}=-\frac{2r^2\rho'}{\Sigma\Delta},\nonumber\\
&&G_{t\phi}=\frac{2a\sin^2\theta\left[(r^2+a^2)(a^2\cos^2\theta-r^2)+2r^2\rho\right]\rho'}{\Sigma^3}+\frac{a^2r\sin^2\theta (r^2+a^2)\rho''}{\Sigma^2},\\ \nonumber
&&G_{\theta\theta}=-\frac{2a^2\cos^2\theta\rho'}{\Sigma}-r\rho'',\\ \nonumber
&&G_{\phi\phi}=-\frac{a^2\sin^2\theta\left[(r^2+a^2)(a^2+(2r^2+a^2)\cos2\theta)+2r^3\sin^2\theta\rho\right]\rho'}{\Sigma^3}-\frac{r\sin^2\theta (r^2+a^2)^2\rho''}{\Sigma^2}.\nonumber
\end{eqnarray}
We rewrite the spacetime metric (\ref{2.16}) in the form
\begin{eqnarray}\label{2.16.1}
ds^2=\frac{\Sigma}{\Delta}dr^2+\Sigma d\theta^2-\frac{\Delta}{\Sigma}\left(dt-a\sin^2\theta d\phi\right)^2 +\frac{\sin^2\theta}{\Sigma}\left[(r^2+a^2)d\phi-adt\right]^2
\end{eqnarray}
giving the standard orthonormal basis of the rotating spacetime metric \cite{Mis-Tho-Whe:1973:Gra:}
\begin{eqnarray}\label{basis}
&&e_t^\mu=\frac{1}{\sqrt{\Sigma\Delta}}\left(r^2+a^2,0,0,a\right), \quad
e_r^\mu=\sqrt{\frac{\Delta}{\Sigma}}\left(0,1,0,0\right),\nonumber\\
&&e_\theta^\mu=\frac{1}{\sqrt{\Sigma}}\left(0,0,1,0\right), \quad
e_\phi^\mu=-\frac{1}{\sqrt{\Sigma\sin^2\theta}}\left(a\sin^2\theta,0,0,1\right),
\end{eqnarray}
and finally we obtain the relations between the components of the stress-energy tensor and the Einstein tensor related to the orthonormal basis in the form
\begin{eqnarray}\label{SET}
&&8\pi\epsilon=-e_t^\mu e_t^\nu G_{\mu\nu}, \quad 8\pi p_r=e_r^\mu e_r^\nu G_{\mu\nu}=g^{rr}G_{rr},\nonumber\\
&& 8\pi p_\theta=e_\theta^\mu e_\theta^\nu G_{\mu\nu}=g^{\theta\theta}G_{\theta\theta}, \quad 8\pi p_\phi=-e_\phi^\mu e_\phi^\nu G_{\mu\nu}.
\end{eqnarray}
By inserting the expressions (\ref{Einstein}), (\ref{basis}) and (\ref{SET2}) into (\ref{SET}), we arrive finally at the relations giving the stress-energy tensor of the quinessential field around the rotating black hole:
\begin{eqnarray}\label{SET2}
8\pi\epsilon=-8\pi p_r=\frac{2\rho'r^2}{\Sigma^2}, \quad 8\pi p_\theta=8\pi p_\phi=-8\pi p_r-\frac{\rho''r+2\rho'}{\Sigma}.
\end{eqnarray}

Physical (curvature) singularity of this rotating solution is defined by the point where the Kretschmann scalar $K$ tends to infinity. There is
\begin{eqnarray}\label{singularity2}
K \equiv R^{\mu\nu\varrho\sigma}R_{\mu\nu\varrho\sigma}=\frac{Z(r,a,\theta)}{\Sigma^6},
\end{eqnarray}
where $Z(r,a,\theta)$ is some cumbersome polynomial function of $r$, $\theta$ and $a$. One may see from the expression (\ref{singularity2}) that spacetime metric (\ref{2.16.1}) has singularity along the ring $r^2+a^2\cos^2\theta=0$. This means that a test particle can hit the singularity at $r=0$ when it is moving in the equatorial plane $\theta=\pi/2$ (It is worth noting that $Z(r=0,a,\theta=\pi/2)\neq0$) and that the geometry is not Ricci flat.

We will first discuss the basical properties of the quintessential rotating black hole spacetimes, namely the coordinate singularities governing the loci of the event horizons and the static limit surface determining the boundary of the ergosphere.

\subsection{Event horizons}

The event horizons are defined by the relation $g^{rr}=0$ ($\Delta=0$) that can be expressed in the form of the qubic polynomial equation for the radius of the horizon location
\begin{eqnarray}\label{2.18}
cr^3-r^2+2Mr-a^2=0 .
\end{eqnarray}
For simplicity we can write~(\ref{2.18}) as
\begin{eqnarray}\label{2.19}
c(r-r_-)(r-r_+)(r-r_q)=0 .
\end{eqnarray}
where $r_-$, $r_+$ and $r_q$ represent the inner and outer black hole horizons, and the quintessential cosmological horizon, respectively. Depending on the values of the characteristic parameters $M$, $a$, and $c$ the number of the horizons may decrease from three to one -- the cosmological (quintessential) $r_q$ horizon never vanishes and if only this one horizon remains, the spacetime describes a quintessential rotating naked singularity. Hereafter, we set $M=1$ for simplicity, i.e., we express the spacetime parameters $a$ and $c$ in units of the gravitational mass parameter $M$: $a/M \to a$, $cM \to c$, $r/M \to r$. In the limiting case of the quintessential Schwarzschild black hole, the two horizons, black hole $r_h$ and the quintessential $r_q$ read
\begin{eqnarray}\label{2.20}
r_h = \frac{1-\sqrt{1-8c}}{2c}, \qquad r_q = \frac{1+\sqrt{1-8c}}{2c} ,
\end{eqnarray}
One can see from the expressions (\ref{2.20}) that maximum (critical) value of the quintessence parameter for non-rotating black hole $a=0$ with quintessential field characterized by the parameter $c$ is equal to
\begin{eqnarray}\label{2.21}
 c_{c}(a=0) \equiv \frac{1}{8}.
\end{eqnarray}
We rewrite the loci of the event horizons (\ref{2.18}) of the rotating black holes in terms of the condition for the rotation parameter $a$
\begin{eqnarray}\label{2.22}
a^2 = a_{h}^2(r;c) \equiv cr^3-r^2+2r\ .
\end{eqnarray}
When $c=0$, expression (\ref{2.22}) determines the loci of event horizons of the Kerr black holes. Its local extrema are defined by the expression
\begin{eqnarray}\label{2.22.1}
c=c_e \equiv \frac{2(r-1)}{3r^2}\ .
\end{eqnarray}
The local extrema of the spacetime rotation parameter and quintessential parameters are located at $r=2$ and read
\begin{eqnarray}\label{2.23}
a^2 \leq a_c^2 \equiv \frac{2}{\sqrt{3}}\ , \qquad c \leq c_c \equiv \frac{1}{6}\ ,
\end{eqnarray}
\begin{figure*}[t!.]
\begin{center}
\includegraphics[width=0.47\linewidth]{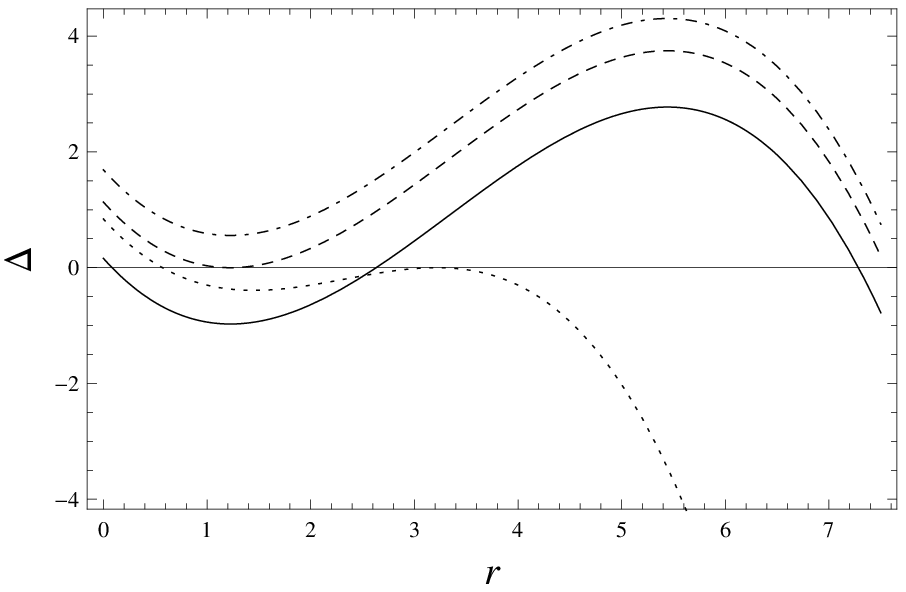}
\includegraphics[width=0.47\linewidth]{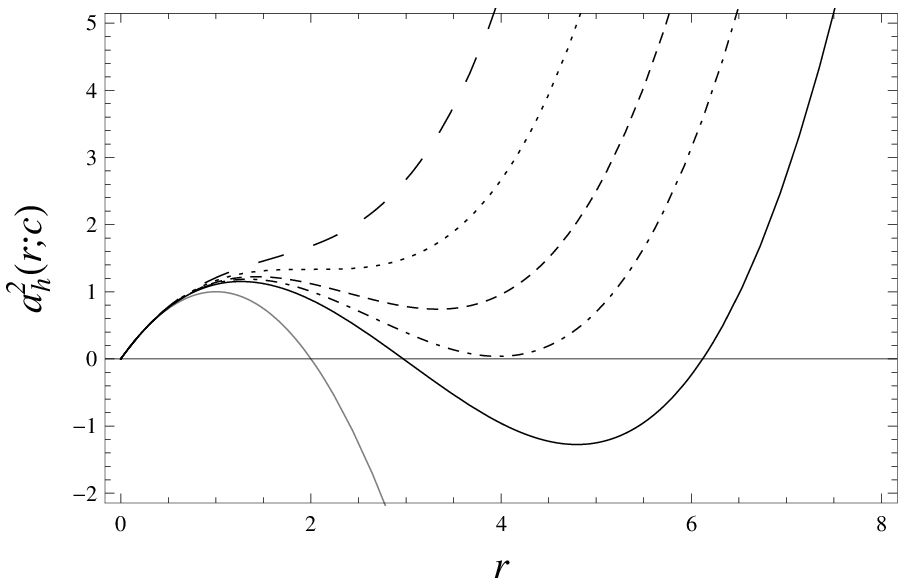}
\end{center}
\caption{\label{fig1} Some possible scenarios can be observed in rotating black hole with quintessential energy with respect to the existence of horizons. \textit{Left panel:} Solid line represents the black hole with inner $r_-$, outer $r_+$ and quintessential $r_q$ horizons, dashed line represents extreme rotating black hole with $r_-=r_+$ and $r_q$ horizons, dotted line represents extreme rotating black hole with $r_-$ and $r_+=r_q$ horizons and finally, dotdashed line represents naked singularity case with only $r_q$ horizon.  \textit{Rigt panel:} Event horizons function $a_{h}^2(r;c)$ of the rotating quintessential black hole for different values of quintessence parameter $c$: grey curve corresponds to the Kerr black hole ($c=0$) horizons. Solid black curve corresponds to $0<c<c_{c}(a=0)\equiv 1/8$. Where black hole can exist for the values of $a^2$ up to local maximum. Dotdashed curve represents the limiting case $c=c_{c}(a=0)$ which is local minimum is located at $a_{h}^2=0$. Dashed curve represents the case $c_{c}(a=0)<c<c_{c}\equiv 1/6$. Dotted curve corresponds to value $c=c_{c}$ which is event horizons and quintessential horizon merge into one. Large dashed line corresponds to the naked singularity case $c>c_{c}$ when only quintessential horizon exists. }
\end{figure*}

In Figure~\ref{fig1} we illustrate the some possible cases of the behavior of the function $\Delta(r;a,c)$ giving dependence of the loci of the inner and outer black hole horizons and the cosmological horizon on the spacetime rotation parameter $a$ for a fixed quintessential parameter $c$.

We see in Figure~\ref{fig1} that the radial dependence of the horizon function $\Delta (r)$ has one local maximum and one local minimum. They can be found from the condition
\begin{eqnarray}\label{2.24}
\frac{d\Delta}{dr}=0\ ,
\end{eqnarray}
that has two solutions
\begin{eqnarray}\label{2.25}
r_{min}=\frac{1-\sqrt{1-6c}}{3c}\ , \qquad r_{max}=\frac{1+\sqrt{1-6c}}{3c}\ ,
\end{eqnarray}
where $r_{min}$ and $r_{max}$ are located between the inner $r_-$ and outer $r_+$ black hole horizons, and the outer black hole $r_+$ and the quintessential cosmological $r_q$ horizons, respectively. From these expressions we obtain the restriction on the values of the quintessential parameter allowing for existence of black hole quintessential spacetimes in the form $c\leq1/6$. The separation of the black hole and naked singularity spacetimes is given by the extremal black hole spacetimes that can be of two types.

\textit{Type I:}
The first type of the extremality occurs when the inner and outer horizons merge into one ($r_-=r_+$) implying the condition
\begin{eqnarray}\label{2.26}
\Delta(r=r_{min})=0\ ,
\end{eqnarray}
giving the first sequence of the extremal black hole spacetimes determined by the relation
\begin{eqnarray}\label{2.27}
a = a_{c1}(r;c) \equiv \sqrt{\frac{2\left(-1+9c+\sqrt{(1-6c)^3}\right)}{27c^2}}\ .
\end{eqnarray}
The inner and outer horizons merge into one namely, the first type of extreme horizon is located at $r_{ext1}=r_{min}$ and quintessential horizon is located at
\begin{eqnarray}\label{2.28}
r_q=\frac{1+2\sqrt{1-6c}}{3c}\ ,
\end{eqnarray}

\textit{Type II:} The second type of the extremality condition corresponds to coincidence of the outer black hole and the quintessential horizons ($r_+=r_q$). Then the extremal black hole spacetimes of the second type have to satisfy the condition
\begin{eqnarray}\label{2.29}
\Delta(r=r_{max})=0\ ,
\end{eqnarray}
implying the second sequence of the extremal black holes determined by the relation
\begin{eqnarray}\label{2.30}
a = a_{c2}(r;c) \equiv \sqrt{\frac{2\left(-1+9c-\sqrt{(1-6c)^3}\right)}{27c^2}}\ .
\end{eqnarray}
The outer and quintessential horizons merge into one namely, the second type of extreme horizon is located at $r_{ext2}=r_{max}$ and inner horizon is located at
\begin{eqnarray}\label{2.31}
r_-=\frac{1-2\sqrt{1-6c}}{3c}\ .
\end{eqnarray}

\textit{Type III:} One can see from the expressions (\ref{2.27}) and (\ref{2.30}) that two types of the extremality conditions coincide when $a_{c1}(c=1/6) = a_{c2}(c=1/6)$ and at this value rotation parameter reaches its maximum
\begin{eqnarray}\label{2.32}
a_{crit} = \frac{2}{\sqrt{3}}.
\end{eqnarray}
In this case all three horizons merge into one. Hereafter, we call this kind of extremal black hole as super extremal black hole.

Thus the black hole spacetimes exist, if the following conditions are satisfied simultaneously
\begin{eqnarray}\label{2.33}
c \leq 1/6 , \quad a \leq a_{c1}(c).
\end{eqnarray}
Otherwise, the quintessential spacetimes are of the naked singularity type.

\begin{figure*}[t!.]
\begin{center}
\includegraphics[width=0.58\textwidth]{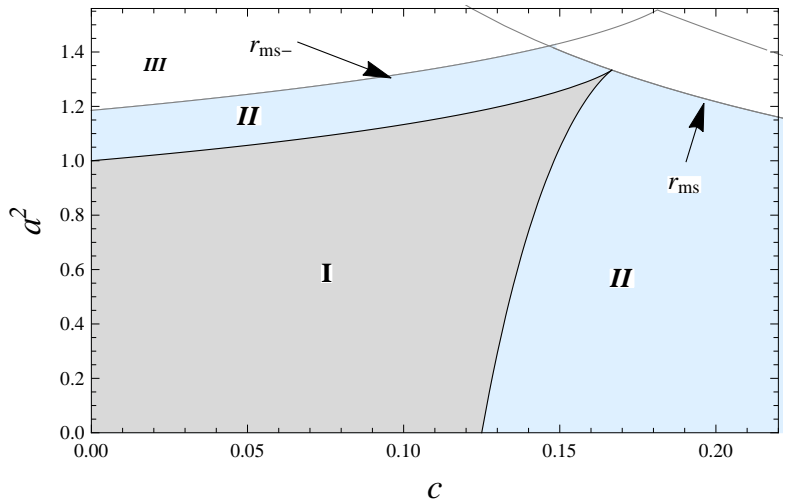}
\end{center}
\caption{\label{fig2} Extremality condition for the rotating black hole with quintessential matter. \textit{Region I}: grey region represents the rotating black hole with all three horizons. \textit{Region II}: light blue region represents naked singularity case (only quintessential horizon) with stable circular orbits. \textit{Region III}: white region represents naked singularity spacetimes that cannot have stable circular orbits. The borders of regions I and II (grey and light blue regions) correspond to the type I ($r_-=r_+$) and type II ($r_+=r_q$) extremality conditions, respectively. The junction of these curves ($c=1/6$ and $a=2/\sqrt{3}$) corresponds to the type III ($r_-=r_+=r_q$) super extreme black hole. Gray curves represent marginally stable orbits.}
\end{figure*}

In Figure~\ref{fig2} we summarize all above results.

\subsection{Ergosphere}

Ergosphere is a region located between the event horizon and the static limit surface defined by the equation $g_{tt}(r_{st},\theta)=0$. The static limit surface is thus governed by the equation
\begin{eqnarray}\label{st-limit}
cr_{st}^3-r_{st}^2+2r_{st}-a^2\cos^2\theta=0 .
\end{eqnarray}
Comparing the equation (\ref{st-limit}) with that of the equation governing the event horizon (\ref{2.18}), one can easily deduce that these two surfaces coincide at the polar caps ($\theta=0,\pi$), if the event horizons exist. In the case of naked singularities the static limit surface is not reaching the symmetry axis, as well known \cite{deFel:1974:ASTRA:,Stu:1980:BULAI:}.

\begin{figure*}[h!.]
\begin{center}
\includegraphics[width=0.32\linewidth]{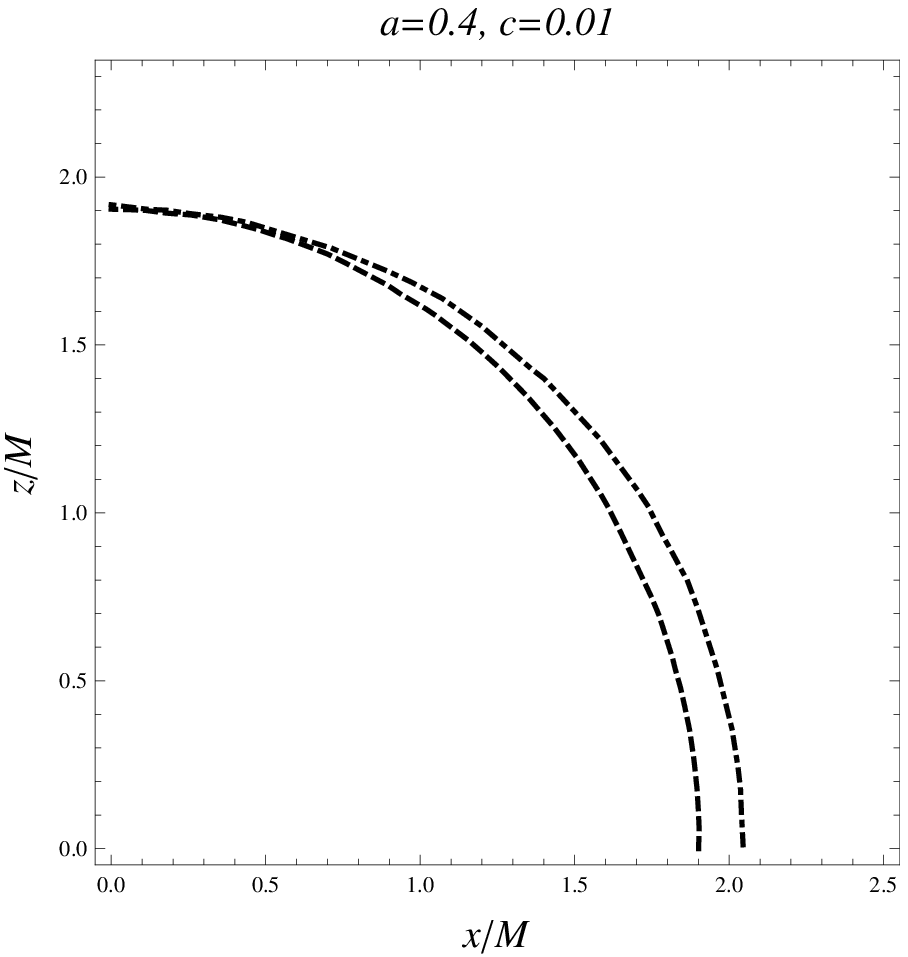}
\includegraphics[width=0.32\linewidth]{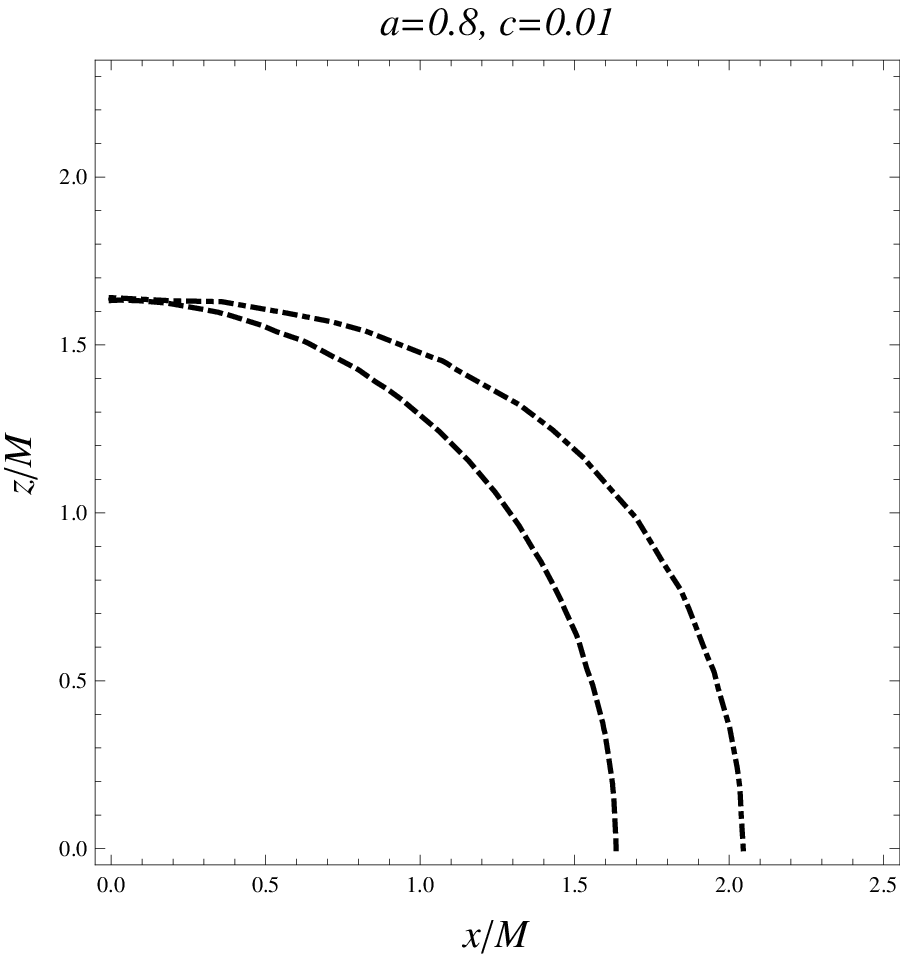}
\includegraphics[width=0.32\linewidth]{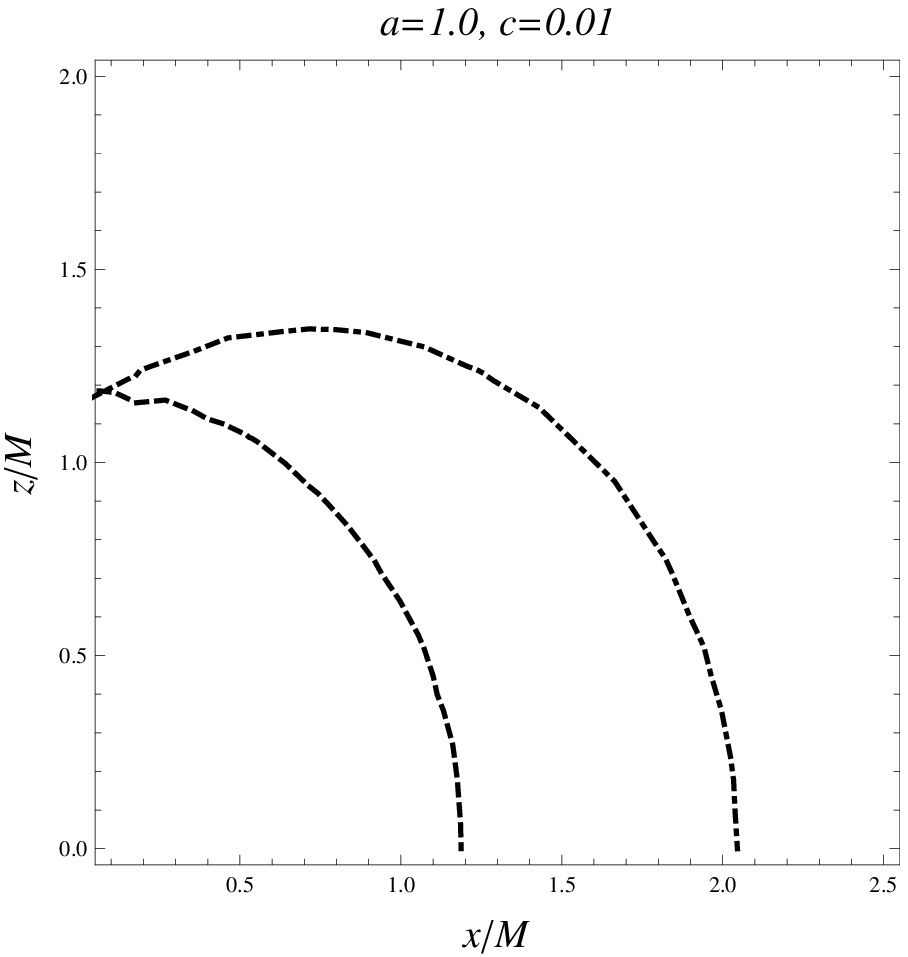}
\includegraphics[width=0.32\linewidth]{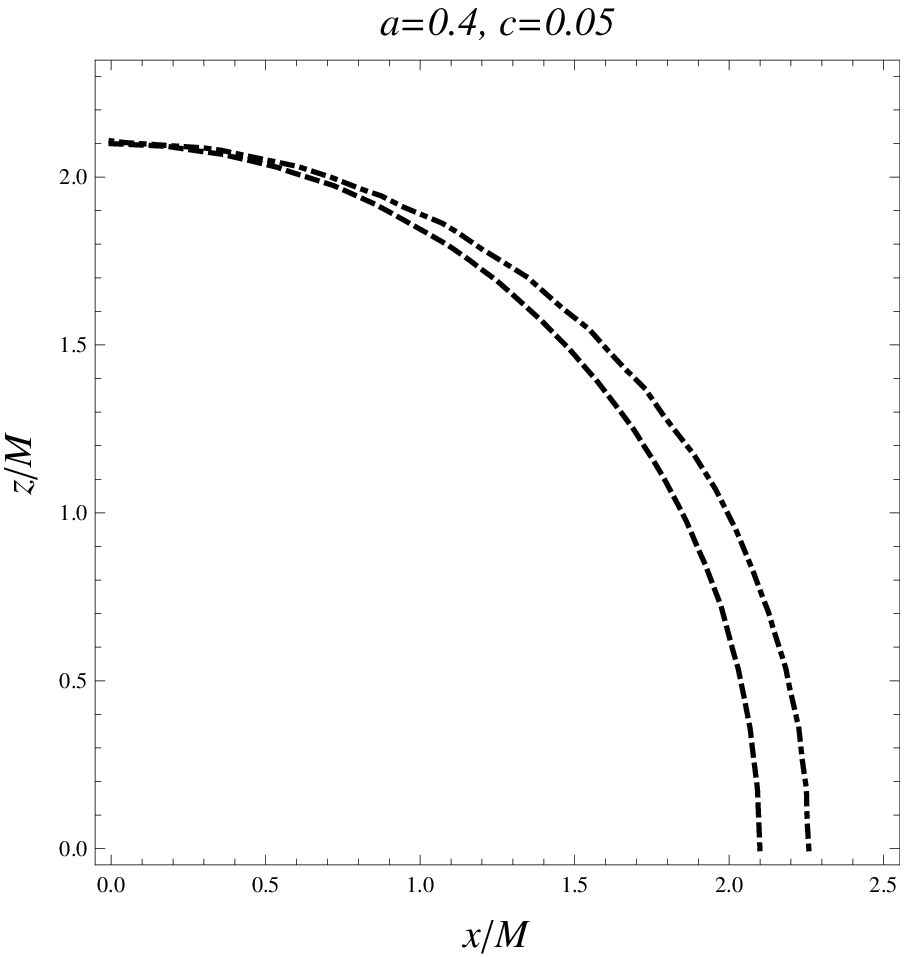}
\includegraphics[width=0.32\linewidth]{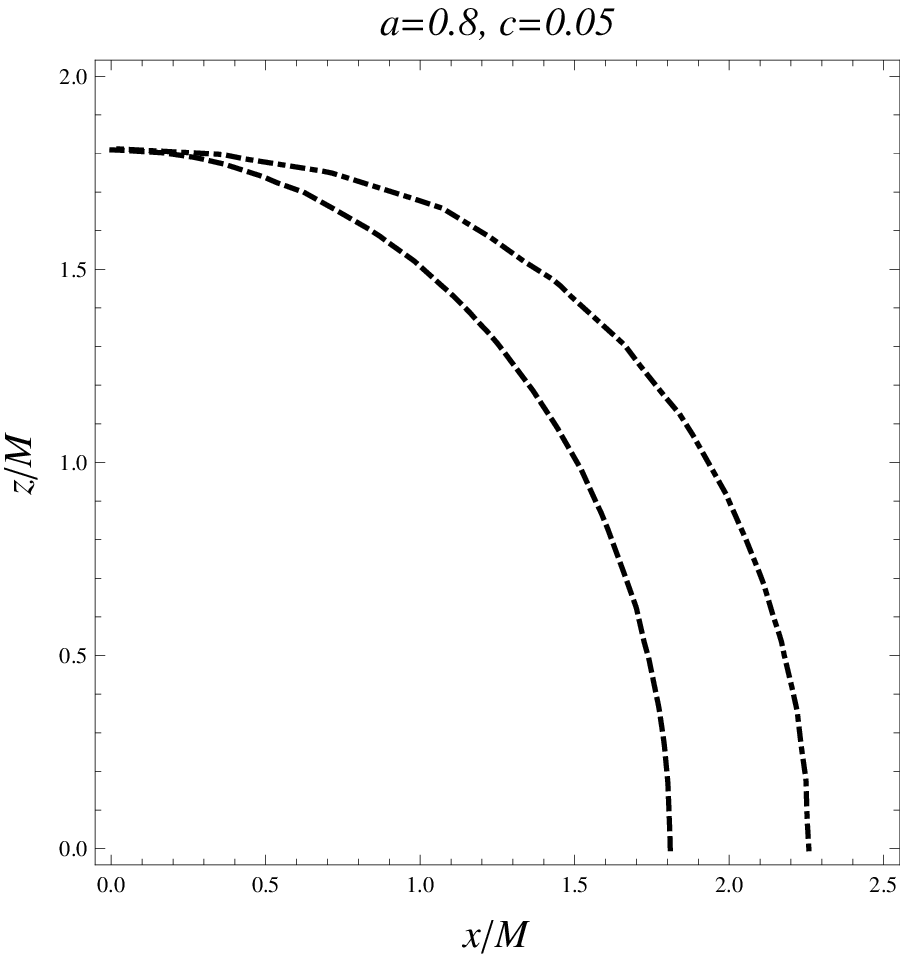}
\includegraphics[width=0.32\linewidth]{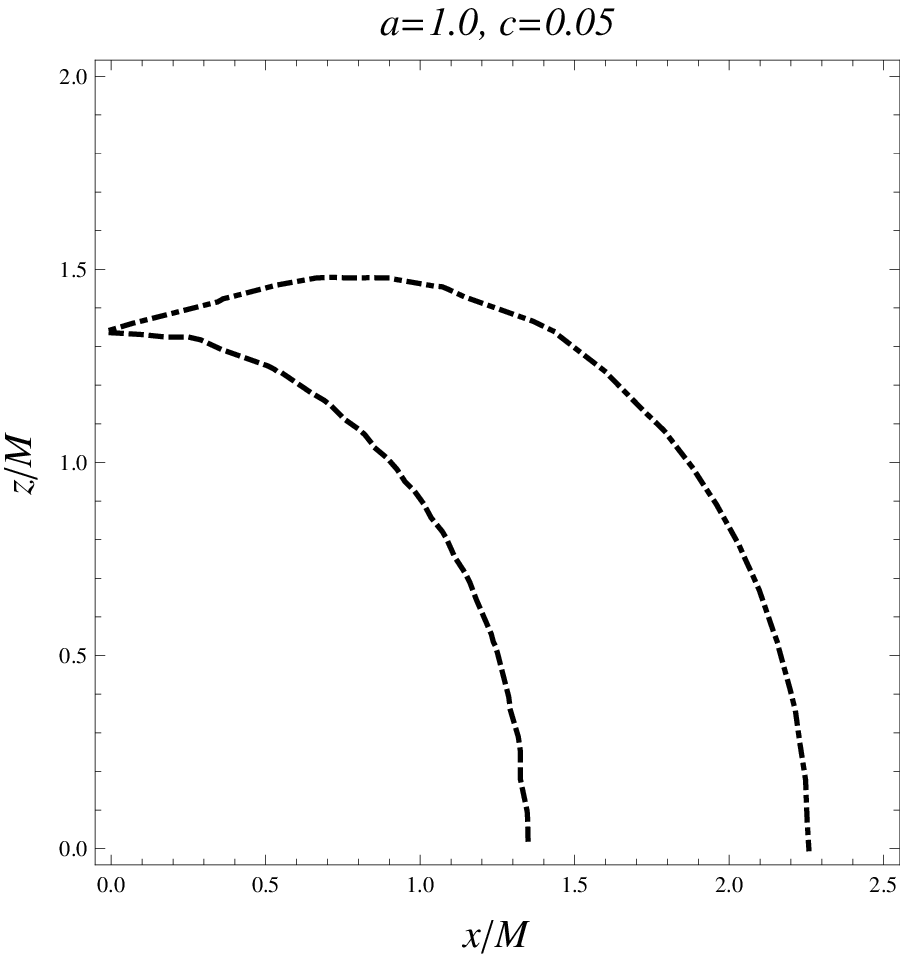}
\end{center}
\caption{\label{fig-ergosphere} Shape and size of ergosphere of the rotating black hole with quintessential matter for different values of the rotation parameter $a$ and quintessence parameter $c$.}
\end{figure*}

Figure~\ref{fig-ergosphere} illustrates behavior of the shape and extension of the ergosphere of the quintessential rotating black holes for various values of the rotation parameter $a$ and the quintessential parameter $c$.

\section{Geodesics structure of the rotating black hole with the quintessential parameter $\omega_{q}=-2/3$}

The geodesic motion in the quintessential rotation black hole spacetimes can be, as in the standard Kerr black hole spacetimes, treated by using the Hamiltonian method \cite{Car:1973:BlaHol:}.

\subsection{Separability of variables in the Hamilton-Jacobi equation}

The Hamilton-Jacobi equation reads
\begin{eqnarray}\label{H-J}
g^{\mu\nu}S_{,\mu}S_{,\nu}=-m^2
\end{eqnarray}
where $S_{,\mu}$ denotes the partial derivative of the action $S$ by the coordinate $x^{\mu}$. $m$is the mass of the test particle. Due to the stationarity and axial symmetry of the spacetime, we can introduce two integrals of the motion, energy $E = -p_t$, and axial angular momentum $L = p_{\phi}$.
Then we can write the Hamilton-Jacobi action function $S$ for a test particle following a geodesic of the quintessential rotating black hole spacetime in the following separated form
\begin{eqnarray}\label{H-J function}
S=-\frac{1}{2}m^2\tau-Et+L\phi+S_r(r)+S_\theta(\theta)\ ,
\end{eqnarray}
where $m^2=0, +1$ is considered for null and timelike geodesics, respectively, $\tau$ is the proper time of the test particle with $m^2=+1$; $m$ can be considered as rest energy of the test particle. Contravariant components of the metric tensor of the spacetime (\ref{01}) have the form
\begin{eqnarray}\label{contravariant}
&&g^{tt}=-\frac{(r^2+a^2)^2-a^2\Delta\sin^2\theta}{\Delta\Sigma}, \quad
g^{t\phi}=-\frac{2a\rho r}{\Delta\Sigma},\nonumber\\
&&g^{rr}=\frac{\Delta}{\Sigma}, \quad g^{\theta\theta}=\frac{1}{\Sigma}, \quad
g^{\phi\phi}=\frac{\Delta-a^2\sin^2\theta}{\Delta\Sigma\sin^2\theta} .
\end{eqnarray}
By inserting (\ref{H-J function}) and (\ref{contravariant}) to (\ref{H-J}), we obtain
\begin{eqnarray}\label{05}
-\left[\frac{(r^2+a^2)^2}{\Delta}-a^2\sin^2\theta\right]E^2+\frac{4a\rho r}{\Delta}EL+ \left[\frac{1}{\sin^2\theta}-\frac{a^2}{\Delta}\right]L^2+\Delta S_{,r}^2+S_{,\theta}^2+m^2\Sigma=0\nonumber\\ .
\end{eqnarray}
Simplifying the equation (\ref{05}), introducing separation constant $K$ representing an additional constant of the motion, and a new constant of the motion through the relation $Q = K - (L - aE)$, we arrive to
\begin{eqnarray}\label{06}
&&\Delta\left(\frac{dS}{dr}\right)^2=\frac{R(r)}{\Delta},\\
&&\left(\frac{dS}{d\theta}\right)^2=\Theta(\theta)\ ,
\end{eqnarray}
where
\begin{eqnarray}\label{06}
&&R(r)=\left[(r^2+a^2)E-aL\right]^2-\Delta\left[(aE-L)^2+m^2r^2+Q\right],\\
&&\Theta(\theta)=Q-\left[\frac{L^2}{\sin^2\theta}+a^2\left(m^2-E^2\right)\right]\cos^2\theta\ .
\end{eqnarray}
We can write the Hamilton-Jacobi action (\ref{H-J function}) in terms of these functions as
\begin{eqnarray}\label{H-J function-2}
S=\frac{1}{2}m^2\tau-Et+L\phi+\int^r\frac{\sqrt{R(r)}}{\Delta}+\int^\theta\sqrt{\Theta(\theta)}.
\end{eqnarray}
By following \cite{Mis-Tho-Whe:1973:Gra:}, we can write the equations of the geodesic motion in the form
\begin{eqnarray}\label{07}
&&\Sigma\dot{t}=\frac{r^2+a^2}{\Delta}\left[E(r^2+a^2)-aL\right]-a(aE\sin^2\theta-L)\ ,\\
&&\Sigma\dot{r}=\sqrt{R}\ ,\\
&&\Sigma\dot{\theta}=\sqrt{\Theta}\ ,\\
&&\Sigma\dot{\phi}=\frac{a}{\Delta}\left[E(r^2+a^2)-aL\right]-\left(aE-\frac{L}{\sin^2\theta}\right)\ .
\end{eqnarray}
where overdot ($\dot{ }$) stands for the derivative with respect to the proper time $\tau$.

\section{Equatorial circular orbits}

\begin{figure*}[h!.]
\begin{center}
\includegraphics[width=0.45\linewidth]{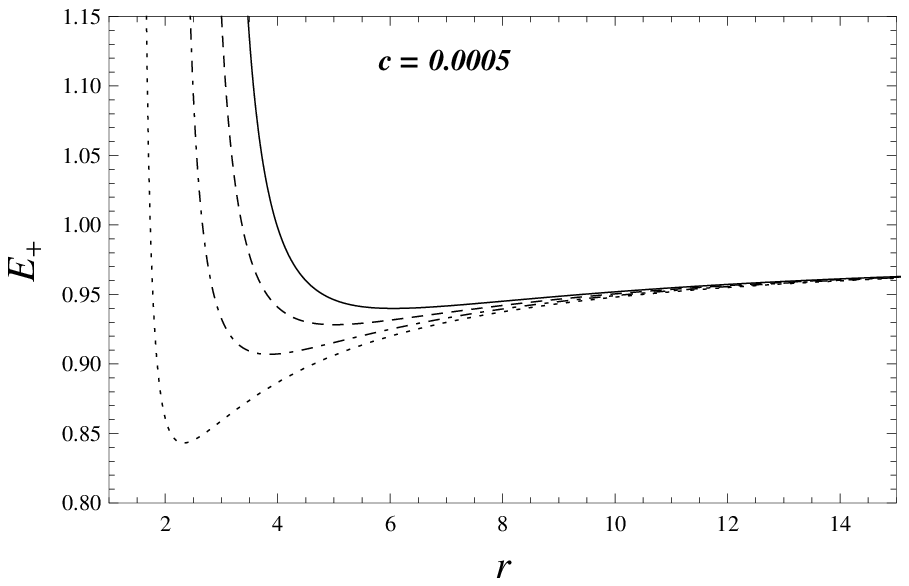}
\includegraphics[width=0.45\linewidth]{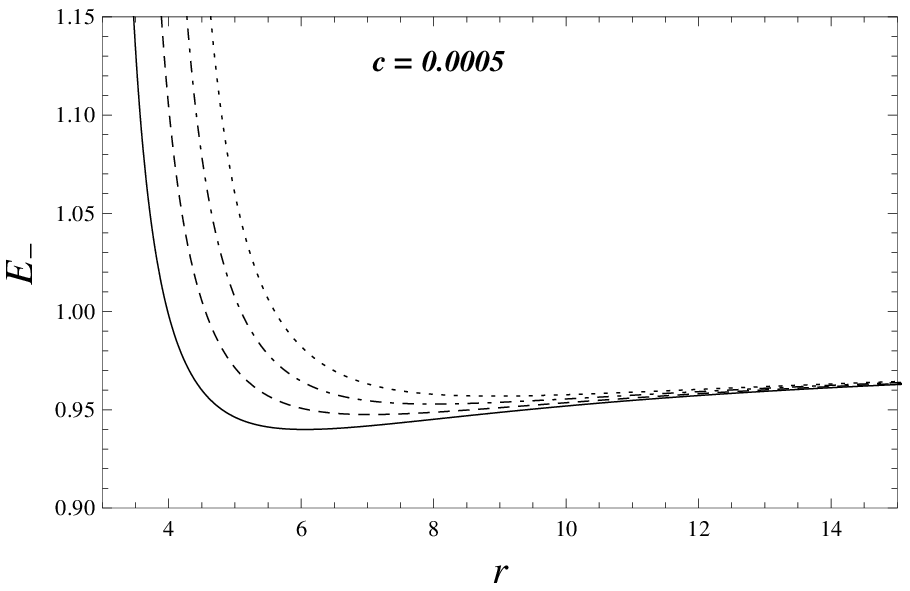}
\includegraphics[width=0.45\linewidth]{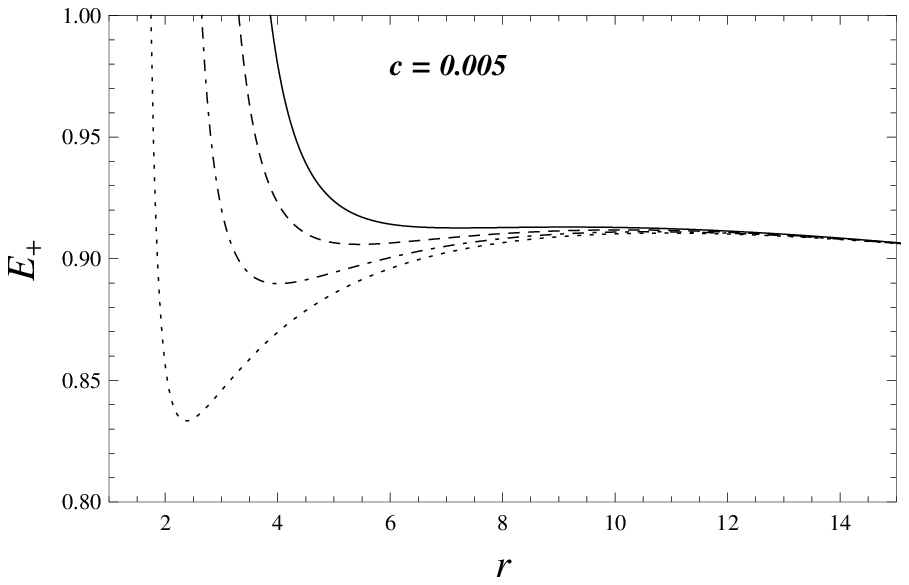}
\includegraphics[width=0.45\linewidth]{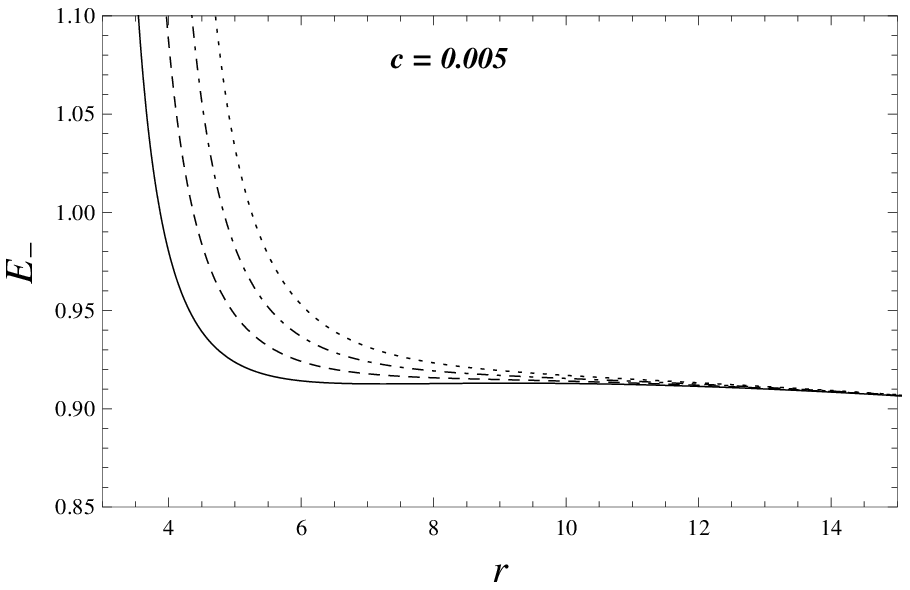}
\end{center}
\caption{\label{fig3} Specific energies of the corotating (left panel) and counterrotating (right panel) particles in the equatorial plane of the rotating black hole with quintessence matter for the values of the rotation parameter $a$: $a=0$ (solid line), $a=0.3$ (dashed line), $a=0.6$ (dotdashed line) and $a=0.9$ (dotted line).}
\end{figure*}

\begin{figure*}[h!.]
\begin{center}
\includegraphics[width=0.45\linewidth]{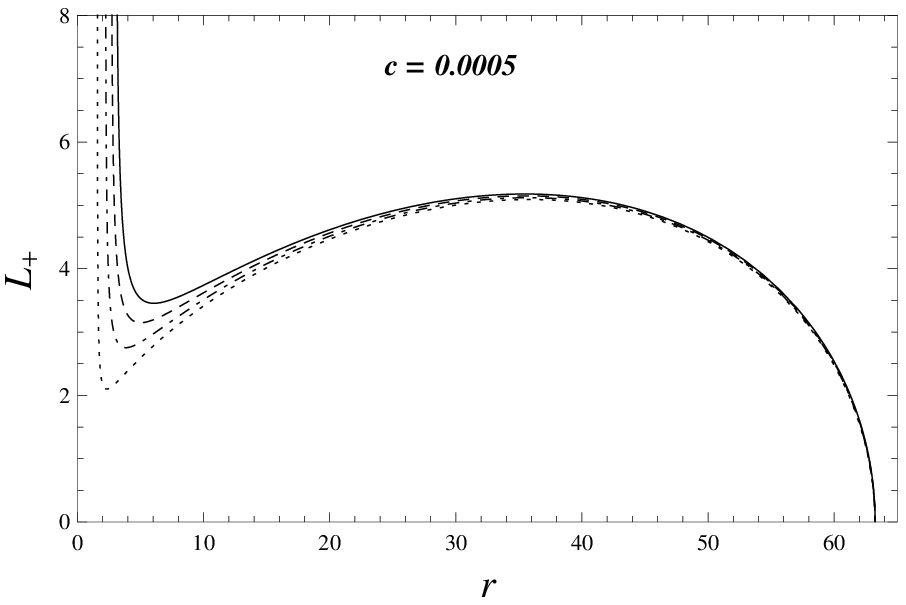}
\includegraphics[width=0.45\linewidth]{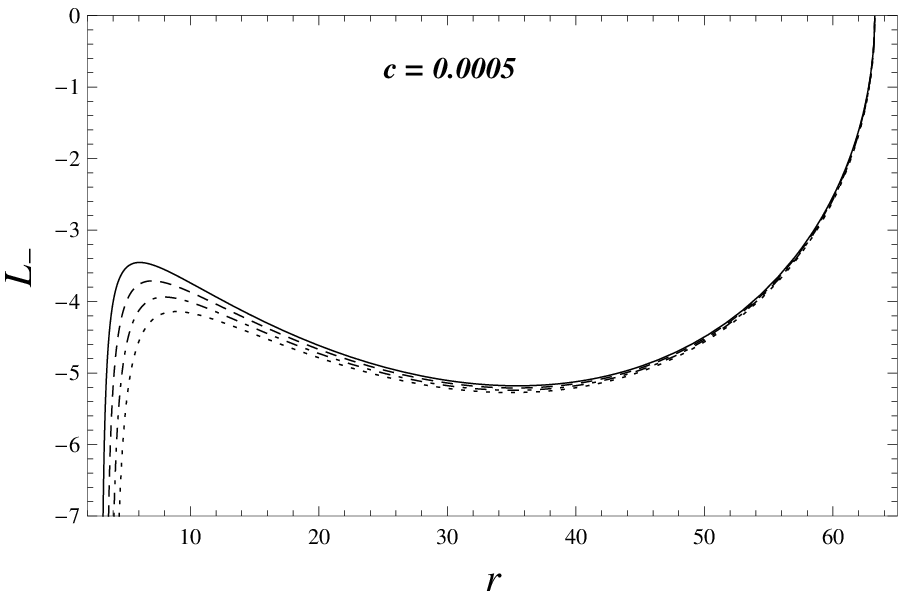}
\includegraphics[width=0.45\linewidth]{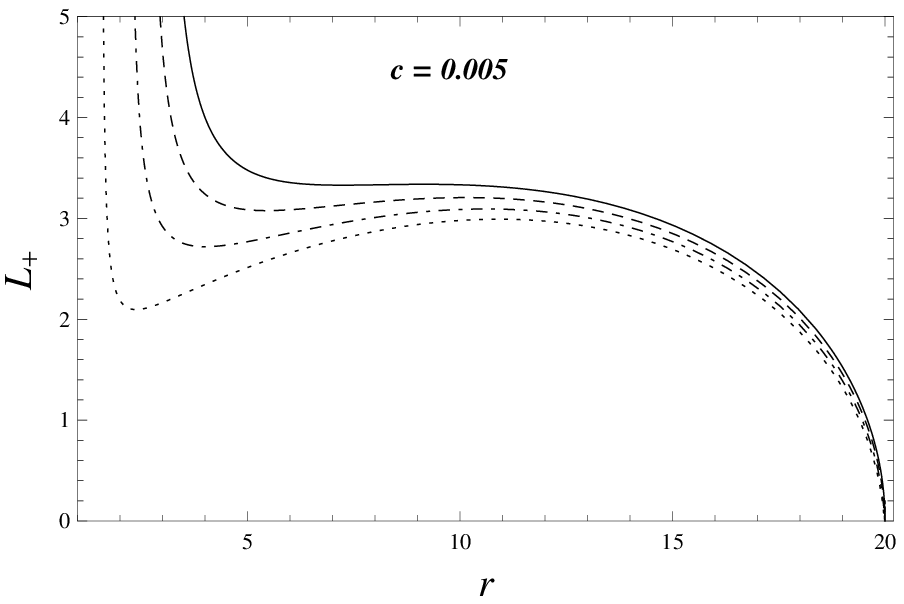}
\includegraphics[width=0.45\linewidth]{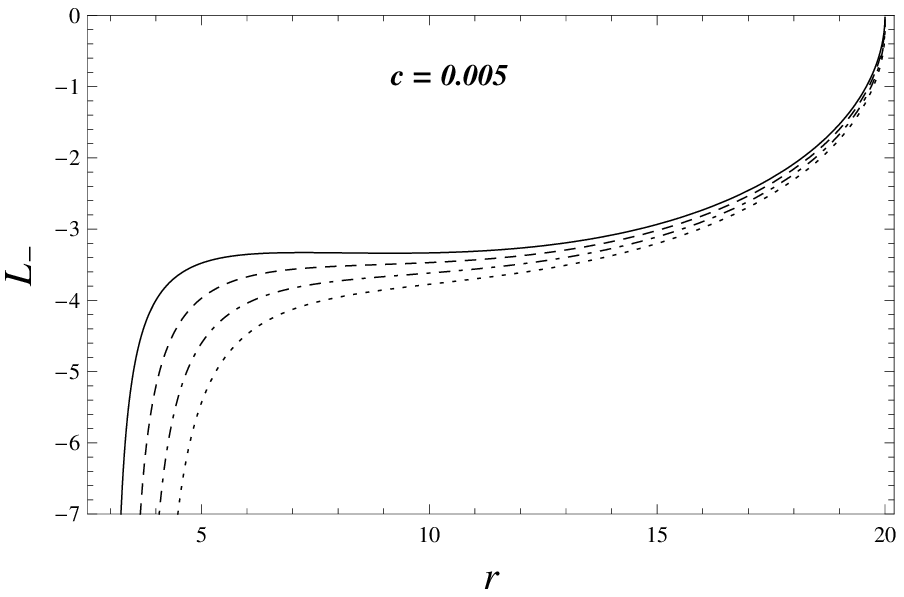}
\end{center}
\caption{\label{fig4} Specific angular momenta of the corotating (left panel) and counterrotating (right panel) particles in the equatorial plane of the rotating black hole with quintessence matter for the values of the rotation parameter $a$: $a=0$ (solid line), $a=0.3$ (dashed line), $a=0.6$ (dotdashed line) and $a=0.9$ (dotted line).}
\end{figure*}

Hereafter, for simplicity we focus on the motion in the equatorial plane $\theta=\pi/2$. In this case, velocity of the particle with respect to $\theta$ axis vanishes $\dot{\theta}=0$ ($\Theta(\pi/2)=0$), consequently $Q=0$. Then, the radial velocity of the particle in the equatorial plane takes the form
\begin{equation}\label{radial}
r^2\dot{r} = \pm R^{1/2}\ ,
\end{equation}
with the function governing the radial motion taking the form
\begin{equation}\label{potential}
R(r)=\left[(r^2+a^2)E-aL\right]^2-\Delta\left[(aE-L)^2+m^2r^2\right] .
\end{equation}
One can find the circular orbits due to the fact that a test particle moving along an equatorial circular orbit has zero radial velocity ($\dot{r}=0$) and zero radial acceleration ($\ddot{r}=0$), i.e.,
\begin{equation}\label{circular}
R(r)=0, \quad \frac{dR}{dr}=0 .
\end{equation}
By solving the above equations simultaneously with respect to $E$ and $L$, we arrive to the following expressions for the specific energy and the specific axial angular momentum of the particle
\begin{equation}\label{energy}
\frac{E_\pm^2}{m^2}=\frac{8\Delta(a^2-\Delta)^2+2r\Delta\Delta'(a^2-\Delta)-a^2r^2\Delta'^2\pm 2\sqrt{2}a\Delta\sqrt{(2a^2-2\Delta+r\Delta')^3}} {r^2\left[16\Delta(\Delta-a^2)+r\Delta'(r\Delta'-8\Delta)\right]},
\end{equation}
\begin{eqnarray}\label{momentum}
\frac{L_\pm^2}{m^2}=&&\frac{8a^2\Delta^3-r^2(r^2+a^2)^2\Delta'^2- 2\Delta^2\left[8a^2(r^2+a^2)+4r^4+a^2r\Delta'\right]}{r^2\left[16\Delta(\Delta-a^2)+r\Delta'(r\Delta'-8\Delta)\right]}\nonumber\\
&&+\frac{2(r^2+a^2) \Delta\left[4a^2(r^2+a^2)+r(3r^2+a^2)\Delta'\right]} {r^2\left[16\Delta(\Delta-a^2)+r\Delta'(r\Delta'-8\Delta)\right]}\nonumber\\
&&\mp\frac{2\sqrt{2}a\Delta\sqrt{(2a^2-2\Delta+r\Delta')^3}\left[(r^2+a^2)(2(r^2+a^2)+r\Delta')-2(2r^2+a^2)\Delta\right]} {r^2\left[16\Delta(\Delta-a^2)+r\Delta'(r\Delta'-8\Delta)\right]},
\end{eqnarray}
where $"'"$ means the derivative with respect to radial coordinate $r$ ($d/dr$). The signs "+" and "-" stand for the corotating and counterrotating particle orbits, respectively. In Figure~\ref{fig3} and Figure~\ref{fig4} specific energies and axial angular momenta of the corotating and counterrotating circular orbits are shown.

\subsection{Static radius giving limit on existence of circular geodesics}

One can see from the expression of the energy of the particle moving along the circular orbit (\ref{energy}) that in order for $E^2$ to be real the expression under square root must be nonnegative, i.e.,
\begin{equation}\label{condition1}
2a^2-2\Delta+r\Delta'\geq0\ .
\end{equation}
Since $\Delta=r^2f(r)+a^2$, this condition implies restriction on the radii of the circular geodesics in the form
\begin{equation}\label{condition1}
r \leq r_s\equiv \left(\frac{2}{c}\right)^{1/2}\ ,
\end{equation}
where $r_s(c)$ is the so called "static radius" where test particles can be located at an unstable equilibrium position, having $L=0$. At this equilibrium point the gravitational attraction of the black hole is just balanced by the cosmic repulsion due the the quintessential field. Interestingly, the static radius $r_s$ does not depend on the rotation parameter $a$, as in the case of the Kerr-de Sitter spacetimes \cite{Stu-Sla:2004:PHYSR4:}.

\subsection{Photon circular geodesics}

Another limit on the existence of the circular geodesics is determined by the photon circular geodesics with radii governed  by the divergence of both the energy and the axial angular momentum of the circular geodesics.
One can see from the expression (\ref{energy}) that the energy of the circular orbits per unit mass goes to infinity ($E_\pm\rightarrow\infty$), if the condition
\begin{equation}\label{photonsphere}
16\Delta(\Delta-a_{ps}^2)+r_{rs}\Delta'(r_{rs}\Delta'-8\Delta)=0\
\end{equation}
is satisfied. By solving this equation with respect to $a^2$ one can find that the radii of photon circular orbits are given by the equation
\begin{equation}\label{photonsphere1}
a^2 = a_{ph}^2(r;c) \equiv \frac{r(cr^2-2r+6)^2}{8(2-cr^2)}\ ,
\end{equation}
where the function $a_{ph}^2(r;q)$ implicitly determines the radii of photon circular orbits.
\begin{figure*}[h!.]
\begin{center}
\includegraphics[width=0.31\linewidth]{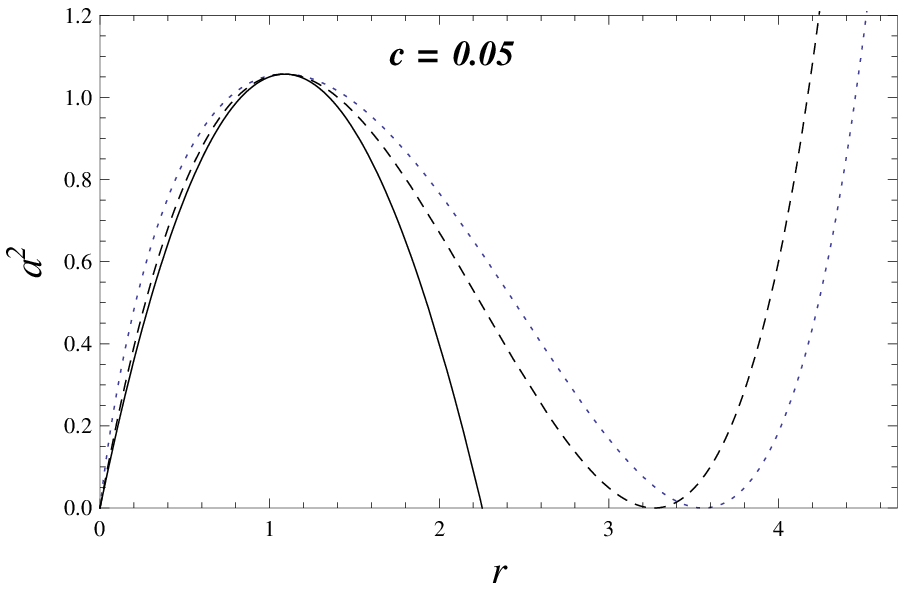}
\includegraphics[width=0.31\linewidth]{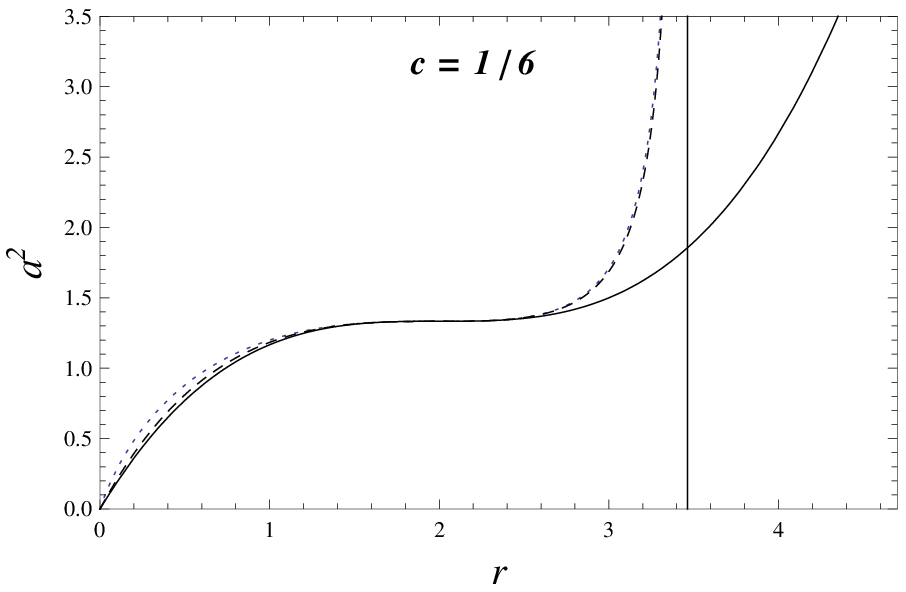}
\includegraphics[width=0.31\linewidth]{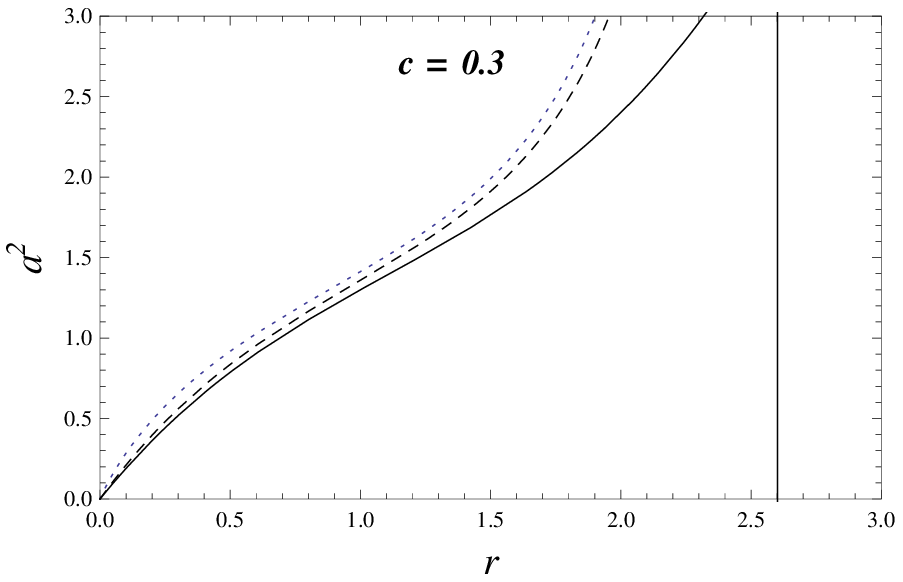}
\end{center}
\caption{\label{fig5} Locations of the horizon (solid curve), photonsphere (dashed curve), marginally bound orbits (dotted curve) and the static radius $r_s=(2/c)^{1/2}$ (solid vertical line). Where black hole (left), extreme black hole (middle) and naked singularity (right) cases are illustrated.}
\end{figure*}
Local maxima of the function $a_{ph}^2(r;q)$ (determined by the equation $\partial_r a_{ps}^2=0$) are located at
\begin{equation}\label{photonsphere2}
r_{e,ph}(c)=\frac{1-\sqrt{1-6c}}{3c}\ .
\end{equation}
Clearly, the photon circular orbits exist in all the quintessential black hole spacetimes. Behavior of the function $a_{ph}^2(r;q)$ for typical values of the quintessential parameter $c$ is illustrated in Figure~\ref{fig5}. We can see that the photon circular orbits exist even in the naked singularity spacetimes -- their radii are shifted towards the center in comparison to the black hole case.


\subsection{Marginally bound circular orbits}

Above the photon circular orbit ($r>r_{ph}$) unbound circular geodesic can exist orbits where a small radial perturbation causes particles following these orbits to fall into the black hole or to escape towards the quintessential horizon. The bound orbits start to exist above the particular radius corresponding to the marginally bound circular orbit (MBCO) $r_{mb}$. If a particle is moving along the orbit with radius $r<r_{mb}$, its energy $E_{\pm}/m^2>1$. The MBCOs are determined by the condition $E_\pm^2/m^2=1$ \cite{Bardeen73} that implies the relation
\begin{eqnarray}\label{mb}
2\Delta^2\left[4\Delta-\Delta'r-8(r^2+a^2)\right]&-&\Delta'^2r^2(r^2+a^2)+ 2\Delta\left[\Delta'r(4r^2+a^2)+4a^2(2r^2+a^2)\right.\nonumber\\
&&\left.\pm\sqrt{2}a\sqrt{(2a^2-2\Delta-\Delta'r)^3}\right]=0 .
\end{eqnarray}
After simplification we obtain the equation for the MBCO radius in the form:
\begin{eqnarray}\label{mb1}
&&2c^3r_0^8-7c^2r_0^7+2c(10c+3)r_0^6-36cr_0^5+4(14c+1)r_0^4-28r_0^3+48r_0^2+a^2r_0(7c^2r_0^4+\nonumber\\
&& 2cr_0^3-4cr_0^2-4r_0-20)\pm2\sqrt{2}(cr_0^3-r_0^2+2r_0-a^2)\sqrt{a^2r_0(2-cr_0^2)^3}=0 .
\end{eqnarray}
The MBCO radius $r_0\equiv r_{mb}$ corresponds to the real smallest root of above equation. By solving (\ref{mb1}) with respect to $a^2$ one can find that the radii of marginally bound orbits are given by the expression
\begin{eqnarray}\label{mb2}
a^2=a_{mb}^2\equiv\frac{2c^2r_0^5+3cr_0^4+8cr_0^3+2r_0^2+8r_0- 2\sqrt{2}(2+cr_0^2)r_0\sqrt{r_0(2+3cr_0^2)}}{2-cr_0^2}.
\end{eqnarray}
The location of the $r_{mb}(a;c)$ is represented in Figure~\ref{fig5} in dependence on rotation parameter $a$ for various values of $c$. In Figure~\ref{fig-mb} the related dependence of the specific angular momentum $\frac{L_{mb}}{m}(a;c)$ is given.

\subsection{Stability of the circular orbits}

\begin{figure*}[h!.]
\begin{center}
\includegraphics[width=0.45\linewidth]{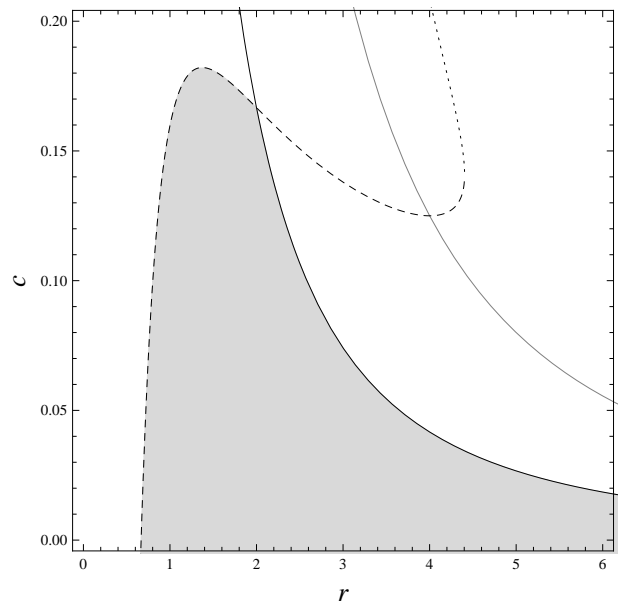}
\end{center}
\caption{\label{fig6} Conditions for the existence of the stable circular orbits. Where dashed and dotted curves represent $c_{ms-}$ and $c_{ms+}$ functions, respectively and moreover, solid black and gray curves correspond to $c_{ms}$ and $c_{s}$, respectively. Shaded region represents the location of stable circular orbits.}
\end{figure*}

The stability of the circular orbits is defined by the inequality
\begin{eqnarray}\label{stab}
\frac{d^2R}{dr^2}\geq0\ .
\end{eqnarray}
The inequality (\ref{stab}) has to be satisfied simultaneously with the conditions of the circular geodesics (\ref{circular}). By inserting the expressions of the specific energy (\ref{energy}) and specific angular momentum (\ref{momentum}) into (\ref{stab}) we transform the inequality into the following form:
\begin{eqnarray}\label{stab1}
\frac{P(r)\pm Q(r)}{R(r)}\geq0\ ,
\end{eqnarray}
where
\begin{eqnarray}\label{stab2}
P(r)&=&r^2(cr^2-2)\left[c^3r^7-5c^2r^6+6c(3c+1)r^5+c(17ca^2-40)r^4\right.\nonumber\\
&&\left.-2(2+c(a^2-30))r^3+(64ca^2-36)r^2+12(a^2-6)r+28a^2\right],\nonumber\\
Q(r)&=&-2\sqrt{2}ar(cr^2-6)(-cr^2+r^2-2r+a^2)\sqrt{r(2-cr^2)^3},\nonumber\\
R(r)&=&(cr^2-2)\left[36r+4(3c+1)r^3-4cr^4+c^2r^5-16a^2+8r^2(ca^2-3)\right].
\end{eqnarray}
The equality in the condition of stability of circular geodesics (\ref{stab1}) corresponds to the marginally stable orbits (innermost stable circular orbit -- ISCO, and outermost stable circular orbit -- OSCO). Loci of these orbits can be given in terms of the rotation parameter $a$ and the quintessential parameter $c$ in an implicit form by the relation
\begin{eqnarray}\label{stab3}
a^2 = a_{ms \pm}^2&=&\frac{-17c^2r^7+3c^2r^6+90c^2r^5-20cr^4-108cr^3+12r^2+56r}{(cr^2-6)^2}\nonumber\\
&&\pm\frac{4r\sqrt{2(2-3cr^2)}\sqrt{(2-cr^2)^3(3c^2r^4-cr^3-12cr^2+6r-4)}}{(cr^2-6)^2}.
\end{eqnarray}
This relation implies that the following conditions related to the spacetime parameter $c$ have to be satisfied in order to have spacetimes allowing for existence of stable circular geodesics:

\begin{itemize}

\item The first condition coincides with the one of the existence of the circular orbits (\ref{condition1}) and it can be expressed in the form $c\leq c_s\equiv2/r^2$.

\item The second condition requires that
\begin{equation}\label{condition2}
c\leq c_{ms}\equiv \frac{2}{3r^2}\ .
\end{equation}
This condition is stronger than the first one. This means that the existence of the circular orbits cannot guarantee their stability.

\item The last condition is implied by the inequality
\begin{equation}\label{condition3}
3c^2r^4-cr^3-12cr^2+6r-4\geq0\ .
\end{equation}
We find that the following inequalities have to be satisfied
\begin{equation}\label{condition3.1}
c\leq c_{ms-}\equiv\frac{12+r-\sqrt{r^2-48r+192}}{6r^2} \quad or \quad
c\geq c_{ms+}\equiv\frac{12+r+\sqrt{r^2-48r+192}}{6r^2}\ .
\end{equation}
\end{itemize}
In Figure~\ref{fig6} we summarize all the restrictions of the quintessential parameter $c$ for the spacetimes allowing for the existence of stable circular geodesics.
One can see from the Figure~\ref{fig6} that stable orbits are located between the curves $c_{ms-}$ and $c_{ms}$. Intersection of $c_{ms-}$ and $c_{ms}$ is at $r=2$ and $c=0.1667$. Moreover, $c_{ms-}$ intersects $c_{s}$ at $r=4$ and $c=0.125$. The quintessence parameter and rotation parameter for existence of the stable circular orbits reaches their maximum (critical) value $c_{crit}$ and $a_{crit}$, respectively at $r=8(3-2\sqrt{2})=1.372583$
\begin{equation}\label{c_crit}
c_{crit}=\frac{3+2\sqrt{2}}{32}=0.182138, \quad a_{crit}=128(-41+29\sqrt{2})=1.249297.
\end{equation}
By comparing the values of the quintessence parameter and rotation parameter for existence of the stable circular orbits (\ref{c_crit}) with (\ref{2.23}) one can deduce that the stable circular orbits can exist also in the quintessential naked singularity spacetimes.

\section{Conclusions}

We have obtained rotating quintessential black hole solution by using the Newman-Janis algorithm. For simplicity we have studied the properties of the rotating quintessential black hole solution with the special value of the quintessential parameter $\omega_q=-2/3$. Using the Einstein field equations, we have determined the stress-energy tensor of the quintessential field around the rotating black holes and naked singularities. We have demonstrated separability and integrability of the equations of geodesic motion, in similarity to the standard Kerr spacetimes. However, in comparison with asymptotically flat Kerr spacetime, in the quintessential rotating spacetimes there are significant differences in the properties of the circular geodesics that are similar to those occuring in the Kerr-de Sitter spacetimes.

Calculations have shown that in the rotating case the upper limit on the quintessence parameter $c$ corresponding to the black hole spacetimes is $c\leq1/6$ that is significantly larger in comparison with the nonrotating case when the limit reads $c\leq1/8$ -- see \cite{You:2015:PHYSR4:}. Moreover, it has been shown that the static radius limiting existence of circular geodesics does not depend on the rotation parameter of the spacetime. The limits on the dimensionless parameters $a$ and $c$ of the quintessential rotating spacetimes are given. The regions of the stable circular geodesics are determined for both the black hole and naked singularity spacetimes.

\begin{acknowledgments}

B.T. and Z.S. would like to express their acknowledgments for the Institutional support of the Faculty of Philosophy and Science of the Silesian University in Opava, the internal student grant of the Silesian University SGS/23/2013 and the Albert Einstein Centre for Gravitation and Astrophysics supported by the Czech Science Foundation grant No.~14-37086G. B.A. acknowledges the Faculty of Philosophy and Science, Silesian University in Opava, Czech Republic and the Goethe University, Frankfurt am Main, Germany for their warm hospitality. The research of B.A. is supported in part by Projects No. F2-FA-F113, No. EF2-FA-0-12477, and No. F2-FA-F029 of the UzAS, and by the ICTP through Grants No. OEA-PRJ-29 and No. OEA-NET-76 and by the Volkswagen Stiftung, Grant No.~86 866.

\end{acknowledgments}

\bibliographystyle{JHEP}
\bibliography{Toshmatov_references}

\providecommand{\href}[2]{#2}\begingroup\raggedright\begin{thebibliography}{10}

\bibitem{Lin:1990:InfCos:}
A.~D. {Linde}, {\it {Particle physics and inflationary cosmology.}},  {\em
  Contemporary Concepts in Physics} {\bf 5} (1990).

\bibitem{Kra-Tur:1995:GENRG2:}
L.~M. {Krauss} and M.~S. {Turner}, {\it {The cosmological constant is back}},
  {\em General Relativity and Gravitation} {\bf 27} (Nov., 1995) 1137--1144,
  [\href{http://arxiv.org/abs/astro-ph/9504003}{{\tt astro-ph/9504003}}].

\bibitem{Ost-Ste:1995:NATURE:}
J.~P. {Ostriker} and P.~J. {Steinhardt}, {\it {The observational case for a
  low-density Universe with a non-zero cosmological constant}},  {\em Nature}
  {\bf 377} (Oct., 1995) 600--602.

\bibitem{Kra:1998:ASTRJ2:}
L.~M. {Krauss}, {\it {The End of the Age Problem, and the Case for a
  Cosmological Constant Revisited}},  {\em The Astrophysical Journal} {\bf 501}
  (July, 1998) 461--466, [\href{http://arxiv.org/abs/astro-ph/9706227}{{\tt
  astro-ph/9706227}}].

\bibitem{Bah-etal:1999:SCIEN:}
N.~A. {Bahcall}, J.~P. {Ostriker}, S.~{Perlmutter}, and P.~J. {Steinhardt},
  {\it {The Cosmic Triangle: Revealing the State of the Universe}},  {\em
  Science} {\bf 284} (May, 1999) 1481,
  [\href{http://arxiv.org/abs/astro-ph/9906463}{{\tt astro-ph/9906463}}].

\bibitem{Cal-Dav-Ste:1998:PHYRL:}
R.~R. {Caldwell}, R.~{Dave}, and P.~J. {Steinhardt}, {\it {Cosmological Imprint
  of an Energy Component with General Equation of State}},  {\em Physical
  Review Letters} {\bf 80} (Feb., 1998) 1582--1585,
  [\href{http://arxiv.org/abs/astro-ph/9708069}{{\tt astro-ph/9708069}}].

\bibitem{ArP-Muk-Ste:2000:PHYRL:}
C.~{Armendariz-Picon}, V.~{Mukhanov}, and P.~J. {Steinhardt}, {\it {Dynamical
  Solution to the Problem of a Small Cosmological Constant and Late-Time Cosmic
  Acceleration}},  {\em Physical Review Letters} {\bf 85} (Nov., 2000) 4438,
  [\href{http://arxiv.org/abs/astro-ph/0004134}{{\tt astro-ph/0004134}}].

\bibitem{Wan-etal:2000:ASTRJ2:}
L.~{Wang}, R.~R. {Caldwell}, J.~P. {Ostriker}, and P.~J. {Steinhardt}, {\it
  {Cosmic Concordance and Quintessence}},  {\em The Astrophysical Journal} {\bf
  530} (Feb., 2000) 17--35, [\href{http://arxiv.org/abs/astro-ph/9901388}{{\tt
  astro-ph/9901388}}].

\bibitem{Ade-etal:2014:ASTRA:}
{Planck Collaboration}, P.~A.~R. {Ade}, N.~{Aghanim}, C.~{Armitage-Caplan},
  M.~{Arnaud}, M.~{Ashdown}, F.~{Atrio-Barandela}, J.~{Aumont},
  C.~{Baccigalupi}, A.~J. {Banday}, and et~al., {\it {Planck 2013 results. XVI.
  Cosmological parameters}},  {\em Astronomy \& Astrophysics} {\bf 571} (Nov.,
  2014) A16, [\href{http://arxiv.org/abs/1303.5076}{{\tt arXiv:1303.5076}}].

\bibitem{Rie-etal:2004:ASTRJ2:}
A.~G. {Riess}, L.-G. {Strolger}, J.~{Tonry}, S.~{Casertano}, H.~C. {Ferguson},
  B.~{Mobasher}, P.~{Challis}, A.~V. {Filippenko}, S.~{Jha}, W.~{Li},
  R.~{Chornock}, R.~P. {Kirshner}, B.~{Leibundgut}, M.~{Dickinson}, M.~{Livio},
  M.~{Giavalisco}, C.~C. {Steidel}, T.~{Ben{\'{\i}}tez}, and Z.~{Tsvetanov},
  {\it {Type Ia Supernova Discoveries at z>1 from the Hubble Space Telescope:
  Evidence for Past Deceleration and Constraints on Dark Energy Evolution}},
  {\em The Astrophysical Journal} {\bf 607} (June, 2004) 665--687,
  [\href{http://arxiv.org/abs/astro-ph/0402512}{{\tt astro-ph/0402512}}].

\bibitem{Spe-etal:2007:ApJSuppl:}
D.~N. {Spergel}, R.~{Bean}, O.~{Dor{\'e}}, M.~R. {Nolta}, C.~L. {Bennett},
  J.~{Dunkley}, G.~{Hinshaw}, N.~{Jarosik}, E.~{Komatsu}, L.~{Page}, H.~V.
  {Peiris}, L.~{Verde}, M.~{Halpern}, R.~S. {Hill}, A.~{Kogut}, M.~{Limon},
  S.~S. {Meyer}, N.~{Odegard}, G.~S. {Tucker}, J.~L. {Weiland}, E.~{Wollack},
  and E.~L. {Wright}, {\it {Three-Year Wilkinson Microwave Anisotropy Probe
  (WMAP) Observations: Implications for Cosmology}},  {\em The Astrophysical
  Journal Supplement Series} {\bf 170} (June, 2007) 377--408,
  [\href{http://arxiv.org/abs/astro-ph/0603449}{{\tt astro-ph/0603449}}].

\bibitem{Cal-Kam:2009:NATURE:CosDarkMat}
R.~{Caldwell} and M.~{Kamionkowski}, {\it {Cosmology: Dark matter and dark
  energy}},  {\em Nature} {\bf 458} (Apr., 2009) 587--589.

\bibitem{Stu:2005:MPLA:}
Z.~{Stuchl{\'{\i}}k}, {\it {Influence of the RELICT Cosmological Constant on
  Accretion Discs}},  {\em Modern Physics Letters A} {\bf 20} (2005) 561--575,
  [\href{http://arxiv.org/abs/0804.2266}{{\tt arXiv:0804.2266}}].

\bibitem{Mis-Tho-Whe:1973:Gra:}
C.~W. {Misner}, K.~S. {Thorne}, and J.~A. {Wheeler}, {\em {Gravitation}}.
\newblock 1973.

\bibitem{Stu:1983:BULAI:}
Z.~{Stuchlik}, {\it {The motion of test particles in black-hole backgrounds
  with non-zero cosmological constant}},  {\em Bulletin of the Astronomical
  Institutes of Czechoslovakia} {\bf 34} (Mar., 1983) 129--149.

\bibitem{Stu:1984:BULAI:}
Z.~{Stuchlik}, {\it {An Einstein-Strauss-de Sitter model of the universe}},
  {\em Bulletin of the Astronomical Institutes of Czechoslovakia} {\bf 35}
  (Aug., 1984) 205--215.

\bibitem{Far-etal:2009:PHYSLETB:}
V.~{Faraoni}, C.~{Gao}, X.~{Chen}, and Y.-G. {Shen}, {\it {What is the fate of
  a black hole embedded in an expanding universe?}},  {\em Physics Letters B}
  {\bf 671} (Jan., 2009) 7--9, [\href{http://arxiv.org/abs/0811.4667}{{\tt
  arXiv:0811.4667}}].

\bibitem{Gre-Lak:2010:PHYSR4:}
C.~{Grenon} and K.~{Lake}, {\it {Generalized Swiss-cheese cosmologies: Mass
  scales}},  {\em Phys. Rev. D} {\bf 81} (Jan., 2010) 023501,
  [\href{http://arxiv.org/abs/0910.0241}{{\tt arXiv:0910.0241}}].

\bibitem{Uza-Ell-Lar:2011:GENRG2:}
J.-P. {Uzan}, G.~F.~R. {Ellis}, and J.~{Larena}, {\it {A two-mass expanding
  exact space-time solution}},  {\em General Relativity and Gravitation} {\bf
  43} (Jan., 2011) 191--205, [\href{http://arxiv.org/abs/1005.1809}{{\tt
  arXiv:1005.1809}}].

\bibitem{Gre-Lak:2011:PHYSR4:}
C.~{Grenon} and K.~{Lake}, {\it {Generalized Swiss-cheese cosmologies. II.
  Spherical dust}},  {\em Phys. Rev. D} {\bf 84} (Oct., 2011) 083506,
  [\href{http://arxiv.org/abs/1108.6320}{{\tt arXiv:1108.6320}}].

\bibitem{Fle-Dup-Uza:2013:PHYSR4:}
P.~{Fleury}, H.~{Dupuy}, and J.-P. {Uzan}, {\it {Can All Cosmological
  Observations Be Accurately Interpreted with a Unique Geometry?}},  {\em
  Physical Review Letters} {\bf 111} (Aug., 2013) 091302,
  [\href{http://arxiv.org/abs/1304.7791}{{\tt arXiv:1304.7791}}].

\bibitem{Far-etal:2014:PHYSR4:}
V.~{Faraoni}, A.~F.~Z. {Moreno}, and A.~{Prain}, {\it {Charged McVittie
  spacetime}},  {\em Phys. Rev. D} {\bf 89} (May, 2014) 103514,
  [\href{http://arxiv.org/abs/1404.3929}{{\tt arXiv:1404.3929}}].

\bibitem{Far:2015:JCAP:}
A.~{Farag Ali}, M.~{Faizal}, and M.~M. {Khalil}, {\it {Short distance physics
  of the inflationary de Sitter universe}},  {\em Journal of Cosmology and
  Astroparticle Physics} {\bf 9} (Sept., 2015) 25,
  [\href{http://arxiv.org/abs/1505.06963}{{\tt arXiv:1505.06963}}].

\bibitem{Far-etal:2015:JCAP:}
V.~{Faraoni}, M.~{Lapierre-L{\'e}onard}, and A.~{Prain}, {\it {Turnaround
  radius in an accelerated universe with quasi-local mass}},  {\em Journal of
  Cosmology and Astroparticle Physics} {\bf 10} (Oct., 2015) 13,
  [\href{http://arxiv.org/abs/1508.01725}{{\tt arXiv:1508.01725}}].

\bibitem{McV:1933:MONRAS:}
G.~C. {McVittie}, {\it {The mass-particle in an expanding universe}},  {\em
  Monthly Notices of the Royal Astronomical Society} {\bf 93} (Mar., 1933)
  325--339.

\bibitem{Nol:1998:PHYSR4}
B.~C. {Nolan}, {\it {A point mass in an isotropic universe: Existence,
  uniqueness, and basic properties}},  {\em Phys. Rev. D} {\bf 58} (Sept.,
  1998) 064006, [\href{http://arxiv.org/abs/gr-qc/9805041}{{\tt
  gr-qc/9805041}}].

\bibitem{Nol:1999:CLAQG:}
B.~C. {Nolan}, {\it {A point mass in an isotropic universe: II. Global
  properties}},  {\em Classical and Quantum Gravity} {\bf 16} (Apr., 1999)
  1227--1254.

\bibitem{Nan-Las-Hob:2012:MONRAS:}
R.~{Nandra}, A.~N. {Lasenby}, and M.~P. {Hobson}, {\it {The effect of a massive
  object on an expanding universe}},  {\em Monthly Notices of the Royal
  Astronomical Society} {\bf 422} (June, 2012) 2931--2944,
  [\href{http://arxiv.org/abs/1104.4447}{{\tt arXiv:1104.4447}}].

\bibitem{Kal-Kle-Mar:2010:PHYSR4:}
N.~{Kaloper}, M.~{Kleban}, and D.~{Martin}, {\it {McVittie's legacy: Black
  holes in an expanding universe}},  {\em Phys. Rev. D} {\bf 81} (May, 2010)
  104044, [\href{http://arxiv.org/abs/1003.4777}{{\tt arXiv:1003.4777}}].

\bibitem{Lak-Abd:2011:PHYSR4:}
K.~{Lake} and M.~{Abdelqader}, {\it {More on McVittie's legacy: A
  Schwarzschild-de Sitter black and white hole embedded in an asymptotically
  {$\Lambda$}CDM cosmology}},  {\em Phys. Rev. D} {\bf 84} (Aug., 2011) 044045,
  [\href{http://arxiv.org/abs/1106.3666}{{\tt arXiv:1106.3666}}].

\bibitem{Sil-Fon-Gua:2013:PHYSR4:}
A.~M. {da Silva}, M.~{Fontanini}, and D.~C. {Guariento}, {\it {How the
  expansion of the Universe determines the causal structure of McVittie
  spacetimes}},  {\em Phys. Rev. D} {\bf 87} (Mar., 2013) 064030,
  [\href{http://arxiv.org/abs/1212.0155}{{\tt arXiv:1212.0155}}].

\bibitem{Nol:2014:CLAQG:}
B.~C. {Nolan}, {\it {Particle and photon orbits in McVittie spacetimes}},  {\em
  Classical and Quantum Gravity} {\bf 31} (Dec., 2014) 235008,
  [\href{http://arxiv.org/abs/1408.0044}{{\tt arXiv:1408.0044}}].

\bibitem{Stu:2005:MODPLA:}
Z.~{Stuchl{\'{\i}}k}, {\it {Influence of the RELICT Cosmological Constant on
  Accretion Discs}},  {\em Modern Physics Letters A} {\bf 20} (2005) 561--575,
  [\href{http://arxiv.org/abs/0804.2266}{{\tt arXiv:0804.2266}}].

\bibitem{Kot:1918:ANNPH2:PhyBasEinsGr}
F.~{Kottler}, {\it {{\"U}ber die physikalischen Grundlagen der Einsteinschen
  Gravitationstheorie}},  {\em Annalen der Physik} {\bf 361} (1918) 401--462.

\bibitem{Stu-Hle:1999:PHYSR4:}
Z.~{Stuchl{\'{\i}}k} and S.~{Hled{\'{\i}}k}, {\it {Some properties of the
  Schwarzschild-de Sitter and Schwarzschild-anti-de Sitter spacetimes}},  {\em
  Phys. Rev. D} {\bf 60} (Aug., 1999) 044006.

\bibitem{Stu:2000:ACTPS2:}
Z.~{Stuchl{\'{\i}}k}, {\it {Spherically symmetric static configurations of
  uniform density in spacetimes with a non-zero cosmological constant}},  {\em
  ACTA PHYSICA SLOVACA} {\bf 50} (Apr., 2000) 219--228.

\bibitem{Boh:2004:GENRG2:}
C.~G. {B{\"o}hmer}, {\it {Eleven Spherically Symmetric Constant Density
  Solutions with Cosmological Constant}},  {\em General Relativity and
  Gravitation} {\bf 36} (May, 2004) 1039--1054,
  [\href{http://arxiv.org/abs/gr-qc/0312027}{{\tt gr-qc/0312027}}].

\bibitem{Car:1973:BlaHol:}
B.~{Carter}, {\it {Black hole equilibrium states.}},  in {\em Black Holes (Les
  Astres Occlus)} (C.~{Dewitt} and B.~S. {Dewitt}, eds.), pp.~57--214, 1973.

\bibitem{Stu-Cal:1991:GENRG2:}
Z.~{Stuchlik} and M.~{Calvani}, {\it {Null geodesics in black hole metrics with
  non-zero cosmological constant}},  {\em General Relativity and Gravitation}
  {\bf 23} (May, 1991) 507--519.

\bibitem{Stu-Hle:2000:CLAQG:}
Z.~{Stuchl{\'{\i}}k} and S.~{Hled{\'{\i}}k}, {\it {Equatorial photon motion in
  the Kerr-Newman spacetimes with a non-zero cosmological constant}},  {\em
  Classical and Quantum Gravity} {\bf 17} (Nov., 2000) 4541--4576,
  [\href{http://arxiv.org/abs/0803.2539}{{\tt arXiv:0803.2539}}].

\bibitem{Lak:2002:PHYSR4:}
K.~{Lake}, {\it {Bending of light and the cosmological constant}},  {\em Phys.
  Rev. D} {\bf 65} (Apr., 2002) 087301,
  [\href{http://arxiv.org/abs/gr-qc/0103057}{{\tt gr-qc/0103057}}].

\bibitem{Bak-etal:2007:CEURJP:}
P.~{Bakala}, P.~{{\v C}erm{\'a}k}, S.~{Hled{\'{\i}}k}, Z.~{Stuchl{\'{\i}}k},
  and K.~{Truparov{\'a}}, {\it {Extreme gravitational lensing in vicinity of
  Schwarzschild-de Sitter black holes}},  {\em Central European Journal of
  Physics} {\bf 5} (Dec., 2007) 599--610,
  [\href{http://arxiv.org/abs/0709.4274}{{\tt arXiv:0709.4274}}].

\bibitem{Ser:2008:PHYSR4:}
M.~{Sereno}, {\it {Influence of the cosmological constant on gravitational
  lensing in small systems}},  {\em Phys. Rev. D} {\bf 77} (Feb., 2008) 043004,
  [\href{http://arxiv.org/abs/0711.1802}{{\tt arXiv:0711.1802}}].

\bibitem{Sch-Zai:2008:0801.3776:CCTimeDelay}
T.~{Sch{\"u}cker} and N.~{Zaimen}, {\it {Cosmological constant and time
  delay}},  {\em Astronomy and Astrophysics} {\bf 484} (June, 2008) 103--106,
  [\href{http://arxiv.org/abs/0801.3776}{{\tt arXiv:0801.3776}}].

\bibitem{Mul:2008:GENRG2:FallSchBH}
T.~{M{\"u}ller}, {\it {Falling into a Schwarzschild black hole. Geometric
  aspects}},  {\em General Relativity and Gravitation} {\bf 40} (Oct., 2008)
  2185--2199.

\bibitem{Vil-etal:2013:ASTSS1:}
J.~R. {Villanueva}, J.~{Saavedra}, M.~{Olivares}, and N.~{Cruz}, {\it {Photons
  motion in charged Anti-de Sitter black holes}},  {\em Astrophysics and Space
  Science} {\bf 344} (Apr., 2013) 437--446.

\bibitem{Stu-Sla:2004:PHYSR4:}
Z.~{Stuchl{\'{\i}}k} and P.~{Slan{\'y}}, {\it {Equatorial circular orbits in
  the Kerr de Sitter spacetimes}},  {\em Phys. Rev. D} {\bf 69} (Mar., 2004)
  064001, [\href{http://arxiv.org/abs/gr-qc/0307049}{{\tt gr-qc/0307049}}].

\bibitem{Kra:2004:CLAQG:}
G.~V. {Kraniotis}, {\it {Precise relativistic orbits in Kerr and Kerr (anti) de
  Sitter spacetimes}},  {\em Classical and Quantum Gravity} {\bf 21} (Oct.,
  2004) 4743--4769, [\href{http://arxiv.org/abs/gr-qc/0405095}{{\tt
  gr-qc/0405095}}].

\bibitem{Kra:2005:DARK:CCPerPrec}
G.~V. {Kraniotis}, {\it {Frame dragging and bending of light in Kerr and Kerr
  (anti) de Sitter spacetimes}},  {\em Classical and Quantum Gravity} {\bf 22}
  (Nov., 2005) 4391--4424, [\href{http://arxiv.org/abs/gr-qc/0507056}{{\tt
  gr-qc/0507056}}].

\bibitem{Kra:2007:CLAQG:Periapsis}
G.~V. {Kraniotis}, {\it {Periapsis and gravitomagnetic precessions of stellar
  orbits in Kerr and Kerr de Sitter black hole spacetimes}},  {\em Classical
  and Quantum Gravity} {\bf 24} (Apr., 2007) 1775--1808,
  [\href{http://arxiv.org/abs/gr-qc/0602056}{{\tt gr-qc/0602056}}].

\bibitem{Kag-Kun-Lam:2006:PHYLB:SolarSdS}
V.~{Kagramanova}, J.~{Kunz}, and C.~{L{\"a}mmerzahl}, {\it {Solar system
  effects in Schwarzschild de Sitter space time}},  {\em Physics Letters B}
  {\bf 634} (Mar., 2006) 465--470,
  [\href{http://arxiv.org/abs/gr-qc/0602002}{{\tt gr-qc/0602002}}].

\bibitem{Ali:2007:PHYSR4:EMPropKadS}
A.~N. {Aliev}, {\it {Electromagnetic properties of Kerr anti-de Sitter black
  holes}},  {\em Phys. Rev. D} {\bf 75} (Apr., 2007) 084041,
  [\href{http://arxiv.org/abs/hep-th/0702129}{{\tt hep-th/0702129}}].

\bibitem{Ior:2009:NEWASTR:CCDGPGrav}
L.~{Iorio}, {\it {Constraining the cosmological constant and the DGP gravity
  with the double pulsar PSR J0737-3039}},  {\em New Astronomy} {\bf 14} (Feb.,
  2009) 196--199, [\href{http://arxiv.org/abs/0808.0256}{{\tt
  arXiv:0808.0256}}].

\bibitem{Hac-etal:2010:PHYSR4:KerrBHCoStr}
E.~{Hackmann}, C.~{L{\"a}mmerzahl}, V.~{Kagramanova}, and J.~{Kunz}, {\it
  {Analytical solution of the geodesic equation in Kerr-(anti-) de Sitter
  space-times}},  {\em Phys. Rev. D} {\bf 81} (Feb., 2010) 044020,
  [\href{http://arxiv.org/abs/1009.6117}{{\tt arXiv:1009.6117}}].

\bibitem{Oli-etal:2011:MODPLA:ChaParRNadS}
M.~{Olivares}, J.~{Saavedra}, C.~{Leiva}, and J.~R. {Villanueva}, {\it {Motion
  of Charged Particles on the REISSNER-NORDSTR{\"O}M - Sitter Black Hole
  Spacetime}},  {\em Modern Physics Letters A} {\bf 26} (2011) 2923--2950,
  [\href{http://arxiv.org/abs/1101.0748}{{\tt arXiv:1101.0748}}].

\bibitem{Kol-Stu:2010:PHYSR4:}
M.~{Kolo{\v s}} and Z.~{Stuchl{\'{\i}}k}, {\it {Current-carrying string loops
  in black-hole spacetimes with a repulsive cosmological constant}},  {\em
  Phys. Rev. D} {\bf 82} (Dec., 2010) 125012,
  [\href{http://arxiv.org/abs/1103.4005}{{\tt arXiv:1103.4005}}].

\bibitem{Stu-Kol:2012:PHYSR4:}
Z.~{Stuchl{\'{\i}}k} and M.~{Kolo{\v s}}, {\it {Acceleration of string loops in
  the Schwarzschild-de Sitter geometry}},  {\em Phys. Rev. D} {\bf 85} (Mar.,
  2012) 065022, [\href{http://arxiv.org/abs/1206.5658}{{\tt arXiv:1206.5658}}].

\bibitem{Gu-Cheng:2007:GENRG2:}
Z.~{Gu} and H.~{Cheng}, {\it {The circular loop equation of a cosmic string in
  Kerr de Sitter spacetimes}},  {\em General Relativity and Gravitation} {\bf
  39} (Jan., 2007) 1--7, [\href{http://arxiv.org/abs/hep-th/0412297}{{\tt
  hep-th/0412297}}].

\bibitem{Wan-Che:2012:PHYLB:}
L.~{Wang} and H.~{Cheng}, {\it {The evolution of circular loops of a cosmic
  string with periodic tension}},  {\em Physics Letters B} {\bf 713} (June,
  2012) 59--62, [\href{http://arxiv.org/abs/1206.2095}{{\tt arXiv:1206.2095}}].

\bibitem{Mul-Asch:2007:CLAQG:}
A.~{M{\"u}ller} and B.~{Aschenbach}, {\it {Non-monotonic orbital velocity
  profiles around rapidly rotating Kerr (anti-)de Sitter black holes}},  {\em
  Classical and Quantum Gravity} {\bf 24} (May, 2007) 2637--2644,
  [\href{http://arxiv.org/abs/0704.3963}{{\tt arXiv:0704.3963}}].

\bibitem{Sla-Stu:2008:CLAQG:}
Z.~{Stuchl{\'{\i}}k} and S.~{Hled{\'{\i}}k}, {\it {Properties of the
  Reissner-Nordstr$\backslash$''om Spacetimes with a Nonzero Cosmological
  Constant}},  {\em ArXiv e-prints} (Mar., 2008)
  [\href{http://arxiv.org/abs/0803.2685}{{\tt arXiv:0803.2685}}].

\bibitem{Stu-Sla-Hle:2000:ASTRA:}
Z.~{Stuchl{\'{\i}}k}, P.~{Slan{\'y}}, and S.~{Hled{\'{\i}}k}, {\it {Equilibrium
  configurations of perfect fluid orbiting Schwarzschild-de Sitter black
  holes}},  {\em Astronomy and Astrophysics} {\bf 363} (Nov., 2000) 425--439.

\bibitem{Sla-Stu:2005:CLAQG:}
P.~{Slan{\'y}} and Z.~{Stuchl{\'{\i}}k}, {\it {Relativistic thick discs in the
  Kerr de Sitter backgrounds}},  {\em Classical and Quantum Gravity} {\bf 22}
  (Sept., 2005) 3623--3651.

\bibitem{Rez-Zan-Fon:2003:ASTRA:}
L.~{Rezzolla}, O.~{Zanotti}, and J.~A. {Font}, {\it {Dynamics of thick discs
  around Schwarzschild-de Sitter black holes}},  {\em Astronomy and
  Astrophysics} {\bf 412} (Dec., 2003) 603--613,
  [\href{http://arxiv.org/abs/gr-qc/0310045}{{\tt gr-qc/0310045}}].

\bibitem{Stu-etal:2005:PHYSR4:}
Z.~{Stuchl{\'{\i}}k}, P.~{Slan{\'y}}, G.~{T{\"o}r{\"o}k}, and M.~A.
  {Abramowicz}, {\it {Aschenbach effect: Unexpected topology changes in the
  motion of particles and fluids orbiting rapidly rotating Kerr black holes}},
  {\em Phys. Rev. D} {\bf 71} (Jan., 2005) 024037,
  [\href{http://arxiv.org/abs/gr-qc/0411091}{{\tt gr-qc/0411091}}].

\bibitem{Asc:2008:CHIAA:}
B.~{Aschenbach}, {\it {Measurement of Mass and Spin of Black Holes with QPOs}},
   {\em Chinese Journal of Astronomy and Astrophysics Supplement} {\bf 8}
  (Oct., 2008) 291--296, [\href{http://arxiv.org/abs/0710.3454}{{\tt
  arXiv:0710.3454}}].

\bibitem{Kuc-Sla-Stu:2011:JCAP:}
H.~{Kuc{\'a}kov{\'a}}, P.~{Slan{\'y}}, and Z.~{Stuchl{\'{\i}}k}, {\it {Toroidal
  configurations of perfect fluid in the Reissner-Nordstr{\"o}m-(anti-)de
  Sitter spacetimes}},  {\em Journal of Cosmology and Astroparticle Physics}
  {\bf 1} (Jan., 2011) 33.

\bibitem{Stu-Kov:2008:INTJMD:}
Z.~{Stuchl{\'{\i}}k} and J.~{Kov{\'a}{\v r}}, {\it {Pseudo-Newtonian
  Gravitational Potential for Schwarzschild-De Sitter Space-Times}},  {\em
  International Journal of Modern Physics D} {\bf 17} (2008) 2089--2105,
  [\href{http://arxiv.org/abs/0803.3641}{{\tt arXiv:0803.3641}}].

\bibitem{Stu-Sla-Kov:2009:CLAQG:}
Z.~{Stuchl{\'{\i}}k}, P.~{Slan{\'y}}, and J.~{Kov{\'a}{\v r}}, {\it
  {Pseudo-Newtonian and general relativistic barotropic tori in
  Schwarzschild-de Sitter spacetimes}},  {\em Classical and Quantum Gravity}
  {\bf 26} (Nov., 2009) 215013, [\href{http://arxiv.org/abs/0910.3184}{{\tt
  arXiv:0910.3184}}].

\bibitem{Stu-Sch:2011:JCAP:}
Z.~{Stuchl{\'{\i}}k} and J.~{Schee}, {\it {Influence of the cosmological
  constant on the motion of Magellanic Clouds in the gravitational field of
  Milky Way}},  {\em Journal of Cosmology and Astroparticle Physics} {\bf 9}
  (Sept., 2011) 18.

\bibitem{Sche-Stu-Pet:2013:JCAP:}
J.~{Schee}, Z.~{Stuchl{\'{\i}}k}, and M.~{Petr{\'a}sek}, {\it {Influence of the
  cosmic repulsion on the MOND model of the Magellanic Cloud motion in the
  field of Milky Way}},  {\em Journal of Cosmology and Astroparticle Physics}
  {\bf 12} (Dec., 2013) 26, [\href{http://arxiv.org/abs/1312.0817}{{\tt
  arXiv:1312.0817}}].

\bibitem{Stu-Sch:2012:INTJMD:}
Z.~{Stuchl{\'{\i}}k} and J.~{Schee}, {\it {Comparison of General Relativistic
  and Pseudo-Newtonian Description of Magellanic-Clouds Motion in the Field of
  Milky way}},  {\em International Journal of Modern Physics D} {\bf 21} (Apr.,
  2012) 50031.

\bibitem{Gim-Hor:2004:hep-th0405019:GodHolo}
E.~G. {Gimon} and P.~{Ho{\v r}ava}, {\it {Over-Rotating Black Holes, Godel
  Holography and the Hypertube}},  {\em ArXiv High Energy Physics - Theory
  e-prints} (May, 2004) [\href{http://arxiv.org/abs/hep-th/0405019}{{\tt
  hep-th/0405019}}].

\bibitem{Gim-Hor:2009:PHYLB:}
E.~G. {Gimon} and P.~{Ho{\v r}ava}, {\it {Astrophysical violations of the Kerr
  bound as a possible signature of string theory}},  {\em Physics Letters B}
  {\bf 672} (Feb., 2009) 299--302, [\href{http://arxiv.org/abs/0706.2873}{{\tt
  arXiv:0706.2873}}].

\bibitem{Stu-Sch:2012:CLAQG:}
Z.~{Stuchl{\'{\i}}k} and J.~{Schee}, {\it {Observational phenomena related to
  primordial Kerr superspinars}},  {\em Classical and Quantum Gravity} {\bf 29}
  (Mar., 2012) 065002.

\bibitem{deFel:1974:ASTRA:}
F.~{de Felice}, {\it {Repulsive Phenomena and Energy Emission in the Field of a
  Naked Singularity}},  {\em Astronomy and Astrophysics,} {\bf 34} (Aug., 1974)
  15.

\bibitem{deFel:1978:NATURE:}
F.~{de Felice}, {\it {Classical instability of a naked singularity}},  {\em
  Nature} {\bf 273} (June, 1978) 429--431.

\bibitem{Stu:1980:BULAI:}
Z.~{Stuchlik}, {\it {Equatorial circular orbits and the motion of the shell of
  dust in the field of a rotating naked singularity}},  {\em Bulletin of the
  Astronomical Institutes of Czechoslovakia} {\bf 31} (1980) 129--144.

\bibitem{Stu:1981:BULAI:}
Z.~{Stuchlik}, {\it {Evolution of Kerr naked singularities}},  {\em Bulletin of
  the Astronomical Institutes of Czechoslovakia} {\bf 32} (1981) 68--72.

\bibitem{Hio-Mae:2009:PHYSR4:}
K.~{Hioki} and K.-I. {Maeda}, {\it {Measurement of the Kerr spin parameter by
  observation of a compact object's shadow}},  {\em Phys. Rev. D} {\bf 80}
  (July, 2009) 024042, [\href{http://arxiv.org/abs/0904.3575}{{\tt
  arXiv:0904.3575}}].

\bibitem{Stu-Sch:2010:CLAQG:}
Z.~{Stuchl{\'{\i}}k} and J.~{Schee}, {\it {Appearance of Keplerian discs
  orbiting Kerr superspinars}},  {\em Classical and Quantum Gravity} {\bf 27}
  (Nov., 2010) 215017, [\href{http://arxiv.org/abs/1101.3569}{{\tt
  arXiv:1101.3569}}].

\bibitem{Stu-Sch:2013:CLAQG:}
Z.~{Stuchl{\'{\i}}k} and J.~{Schee}, {\it {Ultra-high-energy collisions in the
  superspinning Kerr geometry}},  {\em Classical and Quantum Gravity} {\bf 30}
  (Apr., 2013) 075012.

\bibitem{Kis:2002:CLAQG:}
V.~V. {Kiselev}, {\it {Quintessence and black holes}},  {\em Classical and
  Quantum Gravity} {\bf 20} (Mar., 2003) 1187--1197,
  [\href{http://arxiv.org/abs/gr-qc/0210040}{{\tt gr-qc/0210040}}].

\bibitem{Fer-etal:2003:IJTP:}
S.~{Fernando}, S.~{Meadows}, and K.~{Reis}, {\it {Null Trajectories and Bending
  of Light in Charged Black Holes with Quintessence}},  {\em International
  Journal of Theoretical Physics} {\bf 54} (Oct., 2015) 3634--3653,
  [\href{http://arxiv.org/abs/1411.3192}{{\tt arXiv:1411.3192}}].

\bibitem{New-Jan:1965:JMP:}
E.~T. {Newman} and A.~I. {Janis}, {\it {Note on the Kerr Spinning-Particle
  Metric}},  {\em Journal of Mathematical Physics} {\bf 6} (June, 1965)
  915--917.

\bibitem{Azreg-Ainou2014PRD}
M.~{Azreg-A{\"i}nou}, {\it {Generating rotating regular black hole solutions
  without complexification}},  {\em Phys. Rev. D} {\bf 90} (Sept., 2014)
  064041, [\href{http://arxiv.org/abs/1405.2569}{{\tt arXiv:1405.2569}}].

\bibitem{Azreg-Ainou11}
M.~{Azreg-A{\"\i}nou}, {\it {Comment on 'Spinning loop black holes'}},  {\em
  Classical and Quantum Gravity} {\bf 28} (July, 2011) 148001,
  [\href{http://arxiv.org/abs/1106.0970}{{\tt arXiv:1106.0970}}].

\bibitem{Azreg-Ainou14}
M.~{Azreg-A{\"\i}nou}, {\it {From static to rotating to conformal static
  solutions: rotating imperfect fluid wormholes with(out) electric or magnetic
  field}},  {\em European Physical Journal C} {\bf 74} (May, 2014) 2865,
  [\href{http://arxiv.org/abs/1401.4292}{{\tt arXiv:1401.4292}}].

\bibitem{RGTC}
http://www.inp.demokritos.gr/$\sim$sbonano/RGTC/.

\bibitem{Bardeen73}
J.~M. {Bardeen}, {\it {Timelike and null geodesics in the Kerr metric.}},  in
  {\em Black Holes (Les Astres Occlus)} (C.~{Dewitt} and B.~S. {Dewitt}, eds.),
  pp.~215--239, 1973.

\bibitem{You:2015:PHYSR4:}
A.~{Younas}, M.~{Jamil}, S.~{Bahamonde}, and S.~{Hussain}, {\it {Strong
  gravitational lensing by Kiselev black hole}},  {\em Phys. Rev. D} {\bf 92}
  (Oct., 2015) 084042, [\href{http://arxiv.org/abs/1502.01676}{{\tt
  arXiv:1502.01676}}].

\end{thebibliography}\endgroup

\end{document}